\documentclass[twocolumn]{aastex63}
\usepackage{amsmath}
\usepackage{graphicx}
\usepackage{booktabs}
\usepackage{bm}

\newcommand{\lya}{\rm{Ly}\alpha}

\newcommand{\Phibin}{\Phi_{\rm bin}}
\newcommand{\chired}{\chi^{2}_{\rm red}}

\shorttitle{Space Density of Faint Quasars at $z\sim6$}
\shortauthors{Kim et al.}

\begin{document}

\title{The Infrared Medium-deep Survey. IX. \\Discovery of Two New $z\sim6$ Quasars and Space Density down to $M_{1450}\sim-23.5$ mag}

\email{yjkim.ast@gmail.com, myungshin.im@gmail.com}

\author[0000-0003-1647-3286]{Yongjung Kim}
\affiliation{Kavli Institute for Astronomy and Astrophysics, Peking University, Beijing 100871, People's Republic of China}
\affiliation{Department of Astronomy and Atmospheric Sciences, College of Natural Sciences, Kyungpook National University, Daegu 41566, Republic of Korea}

\author[0000-0002-8537-6714]{Myungshin Im}
\affiliation{SNU Astronomy Research Center, Seoul National University, 1 Gwanak-ro, Gwanak-gu, Seoul 08826, Republic of Korea}
\affiliation{Astronomy Program, Department of Physics \& Astronomy, Seoul National University, 1 Gwanak-ro, Gwanak-gu, Seoul 08826, Republic of Korea}

\author[0000-0003-4847-7492]{Yiseul Jeon}
\affiliation{FEROKA Inc., 411,412, 14 Seongsui-ro 10-gil, Seongdong-gu, Seoul 04784, Republic of Korea}

\author[0000-0002-3560-0781]{Minjin Kim}
\affiliation{Department of Astronomy and Atmospheric Sciences, College of Natural Sciences, Kyungpook National University, Daegu 41566, Republic of Korea}

\author[0000-0003-4176-6486]{Linhua Jiang}
\affiliation{Kavli Institute for Astronomy and Astrophysics, Peking University, Beijing 100871, People's Republic of China}
\affiliation{Department of Astronomy, School of Physics, Peking University, Beijing 100871, People's Republic of China}

\author[0000-0002-2188-4832]{Suhyun Shin}
\affiliation{SNU Astronomy Research Center, Seoul National University, 1 Gwanak-ro, Gwanak-gu, Seoul 08826, Republic of Korea}
\affiliation{Astronomy Program, Department of Physics \& Astronomy, Seoul National University, 1 Gwanak-ro, Gwanak-gu, Seoul 08826, Republic of Korea}

\author{Changsu Choi}
\affiliation{Korea Astronomy and Space Science Institute, Daejeon 34055, Republic of Korea}

\author[0000-0003-4738-4251]{Minhee Hyun}
\affiliation{Korea Astronomy and Space Science Institute, Daejeon 34055, Republic of Korea}
\affiliation{SNU Astronomy Research Center, Seoul National University, 1 Gwanak-ro, Gwanak-gu, Seoul 08826, Republic of Korea}

\author[0000-0003-1470-5901]{Hyunsung D. Jun}
\affiliation{SNU Astronomy Research Center, Seoul National University, 1 Gwanak-ro, Gwanak-gu, Seoul 08826, Republic of Korea}

\author[0000-0002-6925-4821]{Dohyeong Kim}
\affiliation{Department of Earth Sciences, Pusan National University, Busan 46241, Republic of Korea}

\author[0000-0001-5120-0158]{Duho Kim}
\affiliation{Korea Astronomy and Space Science Institute, Daejeon 34055, Republic of Korea}

\author[0000-0002-1710-4442]{Jae-Woo Kim}
\affiliation{Korea Astronomy and Space Science Institute, Daejeon 34055, Republic of Korea}

\author[0000-0002-1418-3309]{Ji Hoon Kim}
\affiliation{SNU Astronomy Research Center, Seoul National University, 1 Gwanak-ro, Gwanak-gu, Seoul 08826, Republic of Korea}
\affiliation{Astronomy Program, Department of Physics \& Astronomy, Seoul National University, 1 Gwanak-ro, Gwanak-gu, Seoul 08826, Republic of Korea}

\author[0000-0002-3810-1806]{Bumhyun Lee}
\affiliation{Korea Astronomy and Space Science Institute, Daejeon 34055, Republic of Korea}
\affiliation{Kavli Institute for Astronomy and Astrophysics, Peking University, Beijing 100871, People's Republic of China}

\author[0000-0001-5342-8906]{Seong-Kook Lee}
\affiliation{SNU Astronomy Research Center, Seoul National University, 1 Gwanak-ro, Gwanak-gu, Seoul 08826, Republic of Korea}
\affiliation{Astronomy Program, Department of Physics \& Astronomy, Seoul National University, 1 Gwanak-ro, Gwanak-gu, Seoul 08826, Republic of Korea}

\author[0000-0002-8136-8127]{Juan Molina}
\affiliation{Kavli Institute for Astronomy and Astrophysics, Peking University, Beijing 100871, People's Republic of China}

\author[0000-0002-2548-238X]{Soojong Pak}
\affiliation{School of Space Research, Kyung Hee University, 1732 Deogyeong-daero, Giheung-gu, Yongin-si, Gyeonggi-do 17104, Republic of Korea}

\author[0000-0002-8292-2556]{Won-Kee Park}
\affiliation{Korea Astronomy and Space Science Institute, Daejeon 34055, Republic of Korea}

\author[0000-0002-0992-5742]{Yoon Chan Taak}
\affiliation{Department of Physics and Astronomy, University of California, Los Angeles, CA 90095-1547, USA}
\affiliation{Astronomy Program, Department of Physics \& Astronomy, Seoul National University, 1 Gwanak-ro, Gwanak-gu, Seoul 08826, Republic of Korea}
\affiliation{SNU Astronomy Research Center, Seoul National University, 1 Gwanak-ro, Gwanak-gu, Seoul 08826, Republic of Korea}

\author[0000-0003-0134-8968]{Yongmin Yoon}
\affiliation{Korea Institute for Advanced Study, 85 Hoegi-ro, Dongdaemun-gu, Seoul 02455, Republic of Korea}




\begin{abstract}
We present the result of the Infrared Medium-deep Survey (IMS) $z\sim6$ quasar survey, using the combination of the IMS near-infrared images and the Canada-France-Hawaii Telescope Legacy Survey (CFHTLS) optical images.
The traditional color-selection method results in 25 quasar candidates over $86$ deg$^{2}$.
We introduce the corrected Akaike Information Criterion (AICc) with the high-redshift quasar and late-type star models to prioritize the candidates efficiently.
Among the color-selected candidates, seven plausible candidates finally passed the AICc selection of which three are known quasars at $z\sim6$.
The follow-up spectroscopic observations for the remaining four candidates were carried out, and we confirmed that two out of four are $z\sim6$ quasars.
With this complete sample, we revisited the quasar space density at $z\sim6$ down to $M_{1450}\sim-23.5$ mag.
Our result supports the low quasar space density at the luminosity where the quasar's ultraviolet ionizing emissivity peaks, favoring a minor contribution of quasars to the cosmic reionization.
\end{abstract}



\section{INTRODUCTION \label{sec:introduction}}

As to which objects produced a large amount of ultraviolet (UV) photons that could rapidly ionize the neutral hydrogen in the high-redshift universe ($z\gtrsim6$; \citealt{McGreer15}), the role of high-redshift active galactic nuclei (AGNs) has been in debate.
The bright and faint populations have been studied by wide-shallow surveys such as the Sloan Digital Sky Survey (SDSS; \citealt{Fan01,Fan06,Jiang08,Jiang09,Jiang15,Jiang16,Yang19}) and narrow-deep surveys like the Cosmic Assembly Near-IR Deep Extragalactic Legacy Survey (CANDELS; \citealt{Giallongo15,Giallongo19,Parsa18,Grazian20}), respectively.
These surveys, however, have not provided a consensus on the number density of intermediate-luminosity AGNs with $M_{1450}\sim-23$ mag (or faint quasars), which make a major contribution to the quasar UV ionizing emissivity \citep{Kim20}.

In the last decade, there have been various attempts to fill the deficiency of the observed high-redshift faint quasar population based on multi-wavelength surveys.
The early works with one or two faint quasars over small survey areas ($\lesssim10$ deg$^{2}$) showed that the quasar luminosity function (LF) at $z\sim6$ has a break like the LFs at lower redshifts, but the space number density is somewhat low at a magnitude fainter than the break remained uncertain \citep{Willott10b,Kashikawa15,Kim15,Onoue17}.
This implies that the quasars can provide only about 10\% or less of the UV ionizing photons required to fully ionize the intergalactic medium (IGM) at $z\sim6$.

Recently, the Subaru High-z Exploration of Low-Luminosity Quasars (SHELLQs) project based on the Hyper Suprime-Cam Subaru Strategic Program (HSC-SSP; \citealt{Aihara18}) has found several dozens of faint quasars over 900 deg$^{2}$ \citep{Matsuoka16,Matsuoka18a,Matsuoka18b,Matsuoka19a,Matsuoka19b}.
With this sample, \cite{Matsuoka18c} derived the quasar LF down to $M_{1450}=-22$ mag, which is extremely suppressed at $M_{1450}>-24$ mag and implies that quasars play only a minor role ($\sim3\%$) in the cosmic reionization at $z\sim6$.

Such a low space density, however, is still inconsistent with that from the faint X-ray AGNs with $M_{1450}\sim-22$ mag \citep{Giallongo15,Giallongo19}, which is an order of magnitude higher than the results from the above studies.
\cite{Matsuoka18c} explained that this discrepancy is due to dust obscuration in UV against by which the X-ray AGNs are not affected (see also \citealt{Trebitsch19}).
But recently, follow-up spectroscopy reveals that GDS 3073, one of their sample, is identified as an AGN in rest-UV \citep{Grazian20}, implying that the number density from the quasars identified by rest-UV spectroscopy is still high at $M_{1450}\sim-22.5$ mag, although their survey area is very small (0.15 deg$^{2}$).
From a different point of view, there are attempts to explain such a discrepancy with a change in the fraction of AGNs outshining its host galaxy at $M_{1450}\gtrsim-24$ mag \citep{Ni20,Kim21}.

We have been performing an independent survey for finding faint $z\sim6$ quasars with the Infrared Medium-deep Survey (IMS; M. Im et al. 2022, in preparation).
This is a moderately deep ($J\sim22.5-23.5$ mag in $5\sigma$ depth) near-infrared (NIR) imaging survey with the Wide Field Camera (WFCam; \citealt{Casali07}) on the United Kingdom InfraRed Telescope (UKIRT), covering $\sim120$ deg${^2}$.
Our main goal is to discover quasars as faint as $M_{1450}\sim-23.5$ mag to figure out the quasar demography in the early universe. 
Combining the NIR data with the optical data from the Canada-France-Hawaii Telescope Legacy Survey (CFHTLS; \citealt{Hudelot12}), we discovered a faint $z\sim6$ quasar and dozens of $z\sim5$ quasars \citep{Kim15,Kim19,Kim20}, and suggested the minor contribution of quasars to the cosmic reionization; quasars provide only $\sim3\%$ of the required photons at $z\sim6$ (up to $15\%$).
In this work, we present an extended result of our $z\sim6$ quasar survey, over the overlap regions between CFHTLS and IMS, covering $\sim86$ deg$^{2}$.
Our main goal is to find quasars as faint as $M_{1450}\sim-23.5$ mag and to examine their space density and implication for the cosmic reionization.

We describe our imaging data in Section \ref{sec:data} and quasar candidate selection in Section \ref{sec:selection}. 
Our follow-up observations in spectroscopy and the discovery of two new $z\sim6$ quasars are described In Section \ref{sec:identify}.
In Section \ref{sec:qlf}, we present the $z\sim6$ quasar space density with our complete sample and discuss the results in Section \ref{sec:discussion}.
Throughout this paper, all the magnitudes are given in AB system and we used the cosmological parameters of $\Omega_{m}=0.3$, $\Omega_{\Lambda}=0.7$, and $H_{0}=70$ km s$^{-1}$ Mpc$^{-1}$.

\section{IMAGING DATA \label{sec:data}}

\begin{deluxetable*}{lccccccccccc}[t]
\tabletypesize{\scriptsize}
\tablecaption{Summary of the Survey Fields\label{tbl:fields}}
\tablehead{
\colhead{Field} & \colhead{R.A.}  & \colhead{Decl.}  &\colhead{Area} & \multicolumn{7}{c}{$5\sigma$ limiting magnitudes (mag) / median seeing ($''$) }\\
& \colhead{(J2000)}  & \colhead{(J2000)}  & \colhead{(deg$^{2}$)} & \colhead{$u^{*}$} & \colhead{$g'$} & \colhead{$r'$} & \colhead{$i'_1$} & \colhead{$i'_2$} & \colhead{$z'$} &  \colhead{$J$}
}
\startdata
XMM-LSS	& 02:22:00 & $-$05:20:00 & 8.7 (5.9/2.8) & 25.7/0.92 & 26.1/0.86 & 25.6/0.71 & 25.3/0.74 & 25.3/0.65 & 24.5/0.71 & 23.4/0.85\\
CFHTLS-W2	& 08:58:00 & $-$03:17:00 & 22.0 (20.4/1.6) & 25.6/0.88 & 26.1/0.80 & 25.5/0.73 & 25.3/0.65 & 25.5/0.61 & 24.2/0.69 & 22.6/0.90 \\
EGS			& 14:18:00 & $+$54:30:00 & 34.4 (29.2/5.2)& 25.7/0.85 & 26.1/0.82 & 25.5/0.73 & 25.2/0.67 & 25.6/0.54 & 24.3/0.64 & 22.6/0.88 \\
SA22		& 22:11:00 & $+$01:50:00 & 21.1 (16.7/4.4) & 25.7/0.82 & 26.2/0.76 & 25.5/0.65 & 25.3/0.64 & 25.5/0.56 & 24.2/0.64 & 23.5/0.82 \\
\enddata
\tablecomments{The coordinates indicate the approximate center of each field. The numbers in parenthesis are the areas observed in $i'_1$ and $i'_2$-bands, respectively. The limiting magnitude is given in a median value for point sources after the PSF correction for an aperture with a diameter of $2\times$FWHM$_{z'}$.}
\end{deluxetable*}

\subsection{IMS \label{sec:ims}}

We use the $J$-band imaging data from the IMS and UKIRT Infrared Deep Sky Survey Deep eXtragalactic Survey (UKIDSS-DXS; \citealt{Lawrence07}), obtained with the Wide Field Camera (WFCam; \citealt{Casali07}) on  the United Kingdom Infrared Telescope (UKIRT).
Each image covers $13\farcm65\times13\farcm65$ area with a pixel scale of $0\farcs2$/pixel after microstepping ($0\farcs4$/pixel in original) .
For simplicity, we hereafter refer to the combination of these two surveys as ``IMS''.

As in \cite{Kim19}, we use the images with rescaled zero-points ($zp$) of 28.0 mag in the Vega system, using the bright coordinate-matched sources from the point-source catalog of the Two Micron All Sky Survey (2MASS; \citealt{Skrutskie06}).
Then we applied the Vega-to-AB correction of 0.938 mag \citep{Hewett06} in the following photometric process.

\subsection{CFHTLS \label{sec:cfhtls}}

In the case of optical data, we used the images from the CFHTLS Wide survey, which were stacked by the TERAPIX processing pipeline\footnote{T0007; \url{http://terapix.iap.fr/eplt/T0007/doc/T0007-doc.html}}.
The images in $u^{*},~g',~r',~i',$ and $z'$ bands were obtained with the MegaCam on the Canada-France Hawaii Telescope (CFHT), and each image covers a $1^{\circ}\times1^{\circ}$ area (hereafter ``tile'') with a pixel scale of $0\farcs186$.
Note that there was a change of $i'$-band filter during the survey, from the filter number of 9701 (or $i'_1$) to 9702 (or $i'_2$).
Unlike stellar sources, it is difficult to constrain well the transition between the $i'_1$ and $i'_2$ magnitudes for high-redshift quasars ($z\sim6$), because their colors dramatically change with respect to their redshifts.
Therefore, we consider the difference between the two $i'$-band filters in the following sections.

For accurate photometry to find faint quasars, we reestimated the $zp$ values of the CFHTLS images.
We first selected the objects that also appear in the point-source catalog of the first data release of the Panoramic Survey Telescope and Rapid Response System (PS1; \citealt{Kaiser02,Chambers16}).
Note that we used the point spread function (PSF) magnitudes from the \texttt{StackObjectThin} table.
The PS1 magnitudes of the selected sources were converted to the MegaCam magnitude system using a conversion relation\footnote{\url{http://www.cadc-ccda.hia-iha.nrc-cnrc.gc.ca/en/megapipe/docs/filt.html}}.
For the $zp$ calculation, we use only the sources within a magnitude range of 17.5 and 21.0 mag to avoid saturated, or low signal-to-noise ratio (S/N) objects that could bias the result.
Then, we compared the PSF magnitudes of the PS1-selected sources with those of the sources extracted from the CFHTLS images using SExtractor \citep{Bertin96} with PSFEx \citep{Bertin11} to determine the $zp$ value of each image in each band.
Most of the offsets from the original $zp$ values provided by \cite{Hudelot12} are less than $0.1$ mag in all bands, but these updated $zp$ values result in the point-source colors that are better in line with the synthetic stellar loci of \cite{Covey07}, including IMS $J$-band magnitudes.
Thus, these reestimated $zp$ values improve the removal of stars during the quasar selection.

\begin{figure}
\centering
\epsscale{1.2}
\plotone{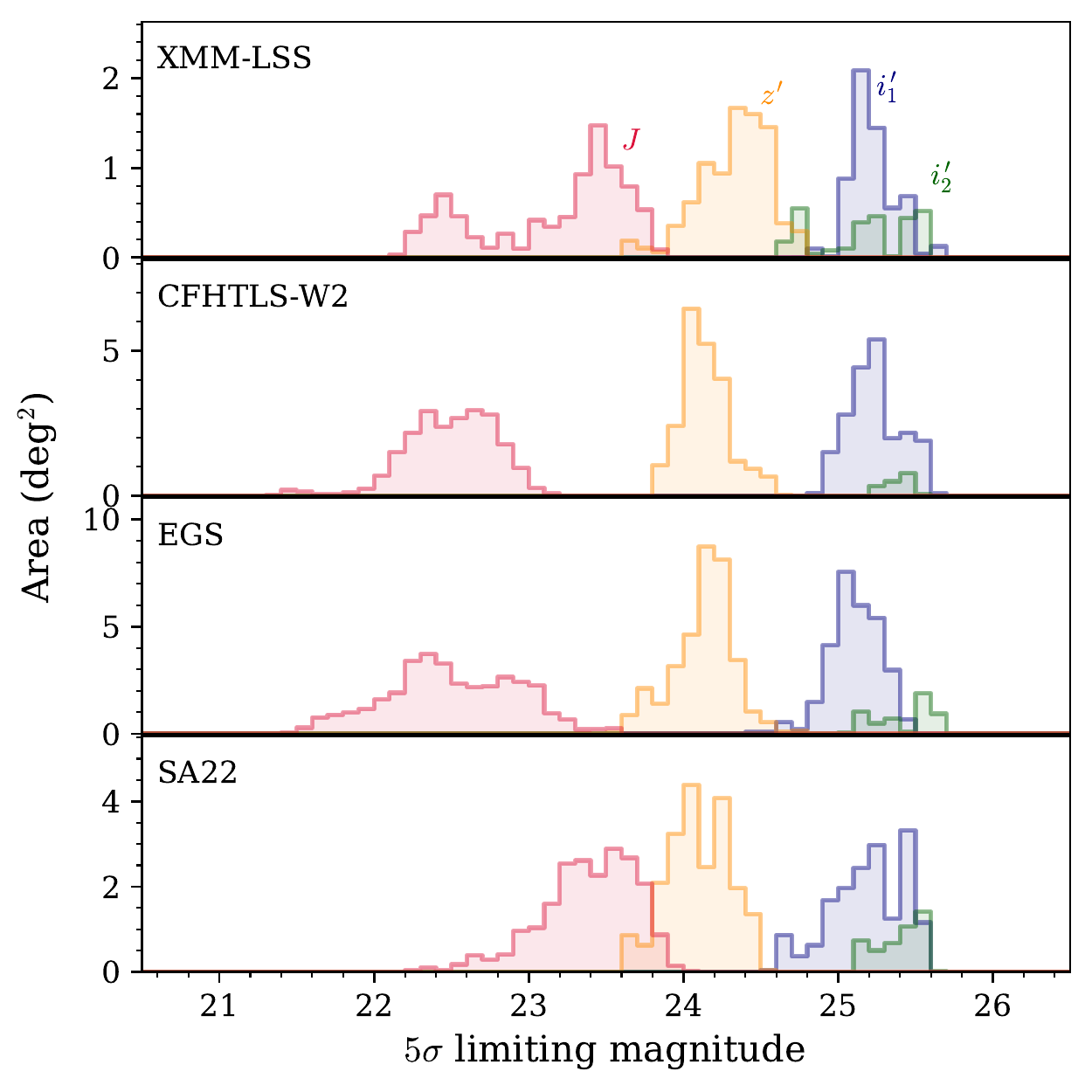}
\caption{
Histogram of the $5\sigma$ limiting magnitudes for a point-source detection of the four survey fields.
Different colors represent the magnitudes in different bands, as marked in the top panel.
\label{fig:histdepth}}
\end{figure}

\subsection{CFHTLS-IMS Overlap\label{sec:overlap}}

There are four extragalactic fields in the CFHTLS-IMS overlap area: the XMM-Large Scale Structure survey region (XMM-LSS), the CFHTLS Wide survey second region (CFHTLS-W2), the Extended Groth Strip (EGS), and the Small Selected Area 22h (SA22).
We resampled the overlap area between CFHTLS and IMS images, using the SWarp software \citep{Bertin10}.
If a region was observed in both the $i'_1$ and $i'_2$ bands, we used the former one.
The four fields cover 8.7, 22.0, 34.4, and 21.1 deg$^{2}$, respectively, and the total sky coverage is 86.2 deg$^{2}$.
The area sizes were calculated from the mosaicked images undersampled to a pixel scale of 1 arcmin/pixel using SWarp\footnote{The area size was updated from that of \citeauthor{Kim20} (\citeyear{Kim20}; 85 deg$^{2}$) with a slight increase.}.
Note that such undersampling is due to the consideration of computing time not only for this area size calculation but also for the survey completeness calculation in Section \ref{sec:comp}.

Using the updated $zp$ values mentioned above, we estimated the limiting magnitudes of each field for point sources, including the PSF correction for an aperture that we used for source extraction (Section \ref{sec:extract}).
In Figure \ref{fig:histdepth}, we show the histogram of the limiting magnitudes in $i'_1$, $i'_2$, $z'$, and $J$-band images.
The detailed information of the four fields including typical image depths is listed in Table \ref{tbl:fields}.
Note that the image depth in a given filter varies between tiles, giving the limiting depth histogram distributions with widths between a few tenths to a couple of magnitudes (Figure \ref{fig:histdepth}).
The optical images in the four fields have homogeneous imaging depths of $u^{*}=25.6$, $g'=26.1$, $r=25.5$, $i'_1=25.3$, $i'_2=25.5$, and $z'=24.2$ mag, with a standard deviation of $\sim0.2$ mag in all bands.
On the other hand, the $J$-band imaging depths show more variations; the depths of the XMM-LSS and SA22 field images are $\sim0.8$ mag deeper than those of the other field images, while portions of the CFHTLS-W2 and EGS fields have shallower depths due to the shorter exposure times. 
We consider this difference when we calculate the survey completeness (Section \ref{sec:comp}).
The median seeing sizes in the ($u^*,~g',~r'~,i'_{1},~i'_{2},~z',~J$) band images are ($0.86,~0.80,~0.71,~0.65,~0.60,~0.68,~0.86$) in units of arcsec, respectively, and those in each field are listed in Table \ref{tbl:fields}.

\subsection{Source Extraction\label{sec:extract}}

With SExtractor, the source detection was performed first in the $z'$-band images at which the Lyman-$\alpha$ $\lambda1216~(\lya)$ emission of a $z\sim6$ quasar is expected to be located.
We set the detection criteria for the SExtractor parameters to $\texttt{DETECT\_MINAREA}=9$ pixels and $\texttt{DETECT\_THRESH}=1.3\,\sigma$, allowing to catalog only the sources with significant ($\gtrsim4\,\sigma$) signals in $z'$-band.
Note that this affects the photometric completeness estimation in Section \ref{sec:comp}.

For the $z'$-band detected sources,  we performed aperture photometry with an aperture of $2\times$FWHM$_{z'}$ diameter,  where FWHM$_{z'}$ is the full-width at half-maximum of point sources in $z'$-band images ($\sim0\farcs7$), by using dual image mode in SExtractor (called forced photometry).
The aperture size is determined to maximize S/N (or \texttt{FLUX}/\texttt{FLUXERR}) of the $z'$-band detection with comparable seeing sizes in the other bands.
The aperture fluxes in each band were converted to the total fluxes by adopting the aperture correction factors derived from bright point-sources in the same field, so that differences in seeing values in different bands are taken care of.
Note that we use aperture instead of PSF because the PSF flux tends to be overestimated if there is no detection when doing forced photometry.

To correct for the galactic extinction (minor in our extragalactic fields; $<0.05$ mag), we used the extinction map of \cite{Schlafly11} with an assumption of $R_V=3.1$ \citep{Cardelli89}.

\section{QUASAR CANDIDATE SELECTION \label{sec:selection}}

\subsection{Point-source Selection\label{sec:pssel}}

\begin{figure}
\centering
\epsscale{1.2}
\plotone{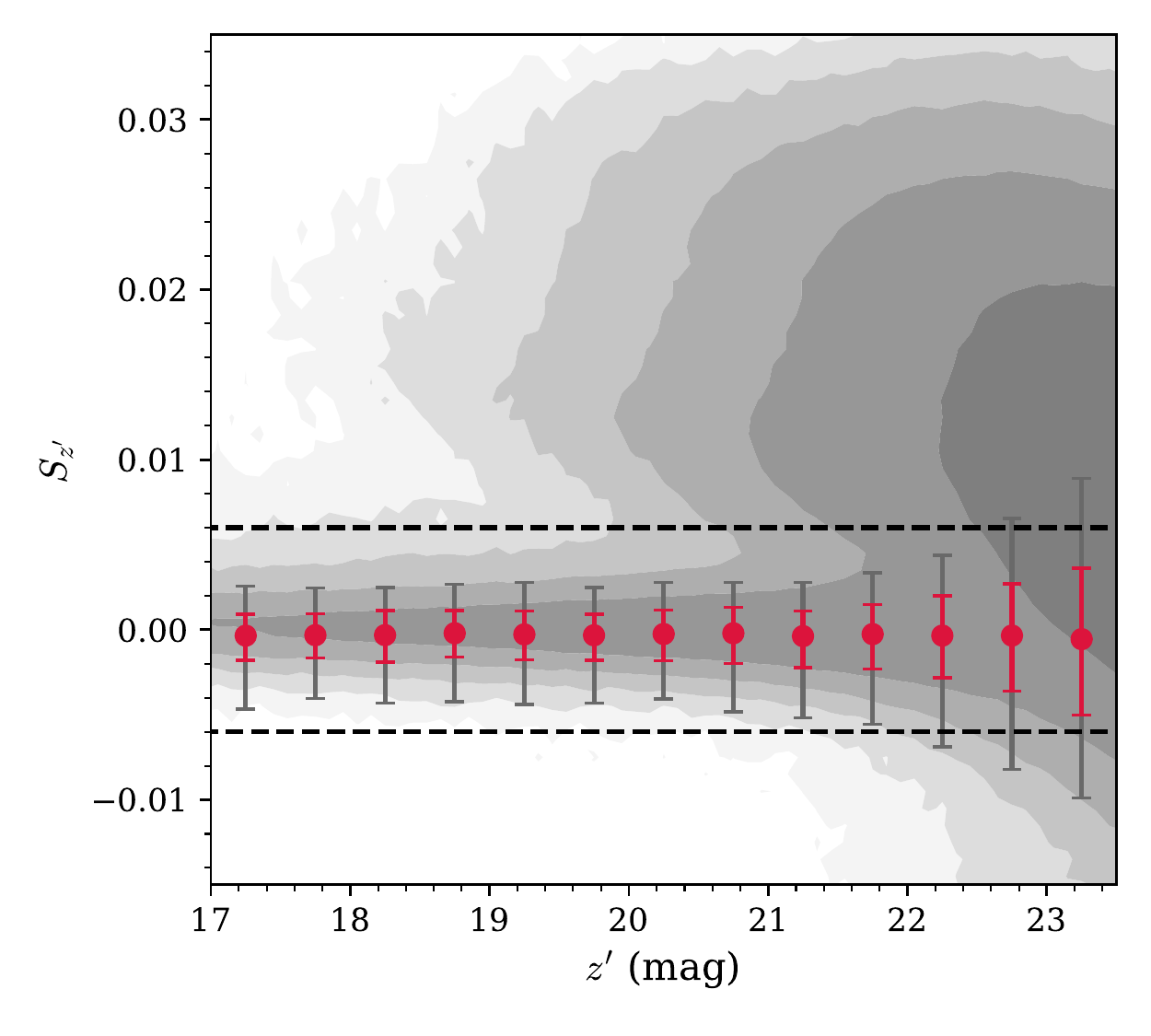}
\caption{
$S_{z'}$ distribution of sources in the four survey fields (gray contours).
The red circles denote the median $S_{z'}$ value of the artificial stars, with the $1\sigma$ (red) and $2\sigma$ (gray) levels.
The dashed lines represent our point-source selection criterion of $-0.006<S_{z'}<0.006$.
\label{fig:spread}}
\end{figure}

Under the imaging resolution of our data, most of the high-redshift quasars with $M_{1450}<-23.5$ mag are classified as point sources ($\gtrsim90\%$; \citealt{Bowler21}).
Previous studies often used the magnitude differences (e.g., PSF magnitude vs aperture magnitude) to avoid the extended-source contamination.
In this work, we adopt the \texttt{SPREAD\_MODEL} parameter, a star-galaxy classifier in SExtractor, which denotes how the source morphology is different from the input point spread function (PSF) model\footnote{\url{https://sextractor.readthedocs.io/en/latest/Model.html}}.
This method offers a better performance to separate point sources from not only the extended sources but also the glitch-like spikes, compared to the previous stellarity index (\texttt{CLASS\_STAR}) in SExtractor, especially at faint magnitudes \citep{Annunziatella13}. 
The \texttt{SPREAD\_MODEL} is defined as

\begin{equation}
\texttt{SPREAD\_MODEL}=\frac{\tilde{\bm{G}}^{T} \bm{W} \bm{p}}{\tilde{\bm{\phi}}^{T} \bm{W} \bm{p}} - \frac{\tilde{\bm{G}}^{T} \bm{W} \tilde{\bm{\phi}}}{\tilde{\bm{\phi}}^{T} \bm{W} \tilde{\bm{\phi}}},
\end{equation}

\noindent where $\bm{p}$ is the image vector centered on the source, and $\bm{W}$ is a weight matrix (diagonal) related to the pixel noises.
$\tilde{\bm{\phi}}$ and $\tilde{\bm{G}}$ represent the point source and the galaxy model vectors at the current position, respectively.
They are based on the resampled local PSF model generated with PSFEx, while the latter ($\tilde{\bm{G}}$) is obtained by convolving an additional circular exponential model.
Since the functional form is normalized by the local PSF model, sources having different PSFs in various fields can be compared to each other.

The average \texttt{SPREAD\_MODEL}  value of point sources is expected to be zero regardless of flux or S/N, but its scatter mildly increases as S/N goes lower.
Therefore, we used the \texttt{SPREAD\_MODEL} value in the $z'$-band ($S_{z'}$) as a reference, considering the high S/N of $z\sim6$ quasars in $z'$-band with $\lya$ emission.
Figure \ref{fig:spread} shows the $S_{z'}$ values of the sources.
There is a clear trend of point sources with $S_{z'}=0$, distinguished from the extended sources ($S_{z'}>0$) and the glitch-like sources ($S_{z'}<0$; unremarkable in this figure with their small numbers).

To test how many point sources can be selected by the arbitrary $S_{z'}$ cut, we performed a simulation by adding artificial stars to the $z'$-band images.
The artificial stars are based on the sampled stars by PSFEx in each image and scaled to match the arbitrary magnitudes that we set for the simulation.
The number of the artificial stars is 100 per 0.5 mag per deg$^{2}$.
Then we repeated the source extraction described in Section \ref{sec:extract}.
The red circles in Figure \ref{fig:spread} show the $S_{z'}$ distribution of the artificial stars, with 1$\sigma$ (68\%) and 2$\sigma$ (95\%) levels, shown as the red and gray colors, respectively.
The widening of the $S_{z'}$-selection range decreases the number of missing point sources, but not surprisingly, the numerous contamination by extended sources also increases.
With several tests, we set the criterion for the point-source selection as $-0.006<S_{z'}<0.006$ (dashed lines) to balance between them down to $z'=23.5$ mag.
With this $S_{z'}$ criterion, 96\% of point sources are recovered at $z'<23.5$ mag, while the rate in the faintest magnitude bin ($23.0\leq z'<23.5$) drops to 83\%.

\begin{figure}
\centering
\epsscale{1.2}
\plotone{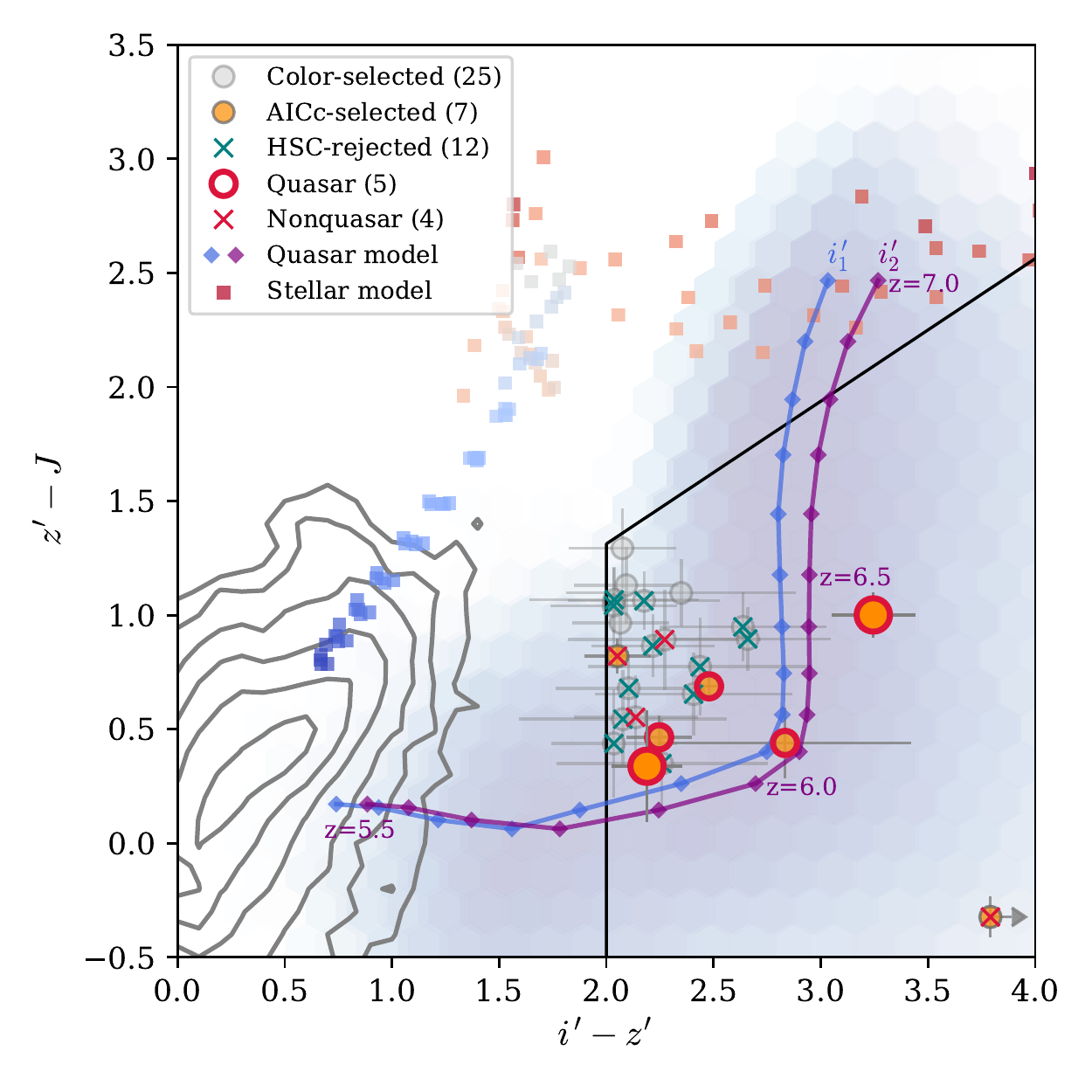}
\caption{
Color-color diagram of point sources in our survey field (gray contours).
The gray circles denote the 25 color-selected candidates with errors, while the arrow indicates the lower limit in color.
The seven AICc-selected candidates with $w_{q}>0.99$ are filled with orange colors.
The rejected candidates after the photometric crossmatch with the HSC data are marked by teal crosses.
The identified quasars and nonquasars are highlighted by the red open circles and crosses, respectively.
The newly discovered quasars are shown as the larger symbols.
The blue and purple diamonds with solid lines are the representative quasar models for $i'_1$ and $i'_2$ magnitudes, respectively, from $z=5.5$ to 7.0 with a step of $\Delta z=0.1$ with $M_{1450}=-24$ mag, $\alpha_{P}=-1.6$, and $\log {\rm EW}=1.542$.
The color distributions of the whole quasar models are shown as the underneath hexagon bins in the same colors.
The late-type star model is shown as the squares, color-coded by the temperature ($T_{\rm eff}=3000$ to 1000 K from blue to red colors).
\label{fig:ccd}}
\end{figure}

\begin{deluxetable*}{lcccccrcl}
\tabletypesize{\scriptsize}
\tablecaption{IMS $z\sim6$ Quasar Candidates \label{tbl:cand}}
\tablehead{
\colhead{ID} & \colhead{R.A. (J2000)} & \colhead{Decl. (J2000)} & \colhead{$i'$} & \colhead{$z'$} & \colhead{$J$} & \colhead{$w_{q}$}  & \colhead{$i'_{\rm HSC}-z'_{\rm HSC}$} & \colhead{Spectroscopy} \\
\colhead{(1)} & \colhead{(2)} & \colhead{(3)} & \colhead{(4)} & \colhead{(5)} & \colhead{(6)} & \colhead{(7)} & \colhead{(8)} & \colhead{(9)} 
}
\startdata
\multicolumn{9}{c}{Color- \& AICc-selected candidates}\\
J140001$+554619^{\rm a}$ & 14:00:01.31 & $+$55:46:19.33 & 25.37$\pm$0.15$^{\rm b}$ & 23.18$\pm$0.07 & 22.85$\pm$0.23 & $>0.99$ & ... & Quasar \\
J140121$+531434$ & 14:01:21.47 & $+$53:14:33.51 & 24.94$\pm$0.14 & 22.89$\pm$0.06 & 22.07$\pm$0.05 & $>0.99$ & ... & Nonquasar \\
J140504$+542435$ & 14:05:03.69 & $+$54:24:34.98 & $>25.70$ & 21.91$\pm$0.03 & 22.23$\pm$0.09 & $>0.99$ & ... & Nonquasar \\
J142952$+544718$ & 14:29:52.18 & $+$54:47:17.68 & 23.97$\pm$0.07 & 21.49$\pm$0.02 & 20.81$\pm$0.04 & $>0.99$ & ... & Quasar \citep{Willott10b} \\
J143055$+531520^{\rm a}$ & 14:30:54.67 & $+$53:15:20.32 & 25.44$\pm$0.19 & 22.19$\pm$0.05 & 21.19$\pm$0.09 & $>0.99$ & ... & Quasar \\
J220418$+011145$ & 22:04:17.93 & $+$01:11:44.77 & 25.15$\pm$0.14 & 22.90$\pm$0.06 & 22.44$\pm$0.07 & $>0.99$ & 2.68$\pm$0.09 & Quasar \citep{Kim15} \\
J221644$-001650$ & 22:16:44.48 & $-$00:16:50.15 & 26.06$\pm$0.58 & 23.23$\pm$0.09 & 22.79$\pm$0.12 & $>0.99$ & 3.34$\pm$0.17 & Quasar \citep{Matsuoka16} \\
\hline
\multicolumn{9}{c}{Color-selected candidates}\\
J022609$-054405$ & 02:26:09.29 & $-$05:44:04.50 & 25.13$\pm$0.28$^{\rm b}$ & 22.91$\pm$0.05 & 22.04$\pm$0.13 & 0.76 & 1.34$\pm$0.04 & ... \\
J084842$-012809$ & 08:48:42.37 & $-$01:28:09.39 & 25.48$\pm$0.27 & 23.44$\pm$0.11 & 23.01$\pm$0.21 & 0.02 & 1.19$\pm$0.10 & ... \\
J085550$-051346$ & 08:55:50.30 & $-$05:13:46.02 & 25.38$\pm$0.27 & 23.25$\pm$0.10 & 22.69$\pm$0.15 & 0.09 & ... & Nonquasar \\
J085756$-050514$ & 08:57:55.94 & $-$05:05:14.21 & 25.22$\pm$0.21 & 23.16$\pm$0.09 & 22.19$\pm$0.09 & 0.01 & ... & ... \\
J090028$-015639$ & 09:00:27.73 & $-$01:56:39.29 & 25.89$\pm$0.51 & 23.46$\pm$0.13 & 22.68$\pm$0.17 & $<0.01$ & 0.92$\pm$0.10 & ... \\
J090554$-052518$ & 09:05:53.65 & $-$05:25:17.94 & 25.75$\pm$0.43 & 23.48$\pm$0.13 & 22.58$\pm$0.18 & $<0.01$ & ... & Nonquasar \\
J141556$+572709$ & 14:15:56.03 & $+$57:27:08.86 & 25.14$\pm$0.23 & 23.05$\pm$0.07 & 21.92$\pm$0.15 & $<0.01$ & ... & ... \\
J141752$+553504$ & 14:17:51.61 & $+$55:35:04.35 & 25.43$\pm$0.24 & 23.35$\pm$0.09 & 22.06$\pm$0.15 & $<0.01$ & ... & ... \\
J143639$+525452$ & 14:36:39.37 & $+$52:54:51.71 & 25.74$\pm$0.52 & 23.39$\pm$0.11 & 22.29$\pm$0.10 & $<0.01$ & ... & ... \\
J220242$+014912$ & 22:02:42.03 & $+$01:49:11.63 & 25.84$\pm$0.45 & 23.43$\pm$0.11 & 22.78$\pm$0.15 & $<0.01$ & 1.20$\pm$0.09 & ... \\
J220350$+012638$ & 22:03:50.19 & $+$01:26:37.69 & 25.55$\pm$0.47 & 23.47$\pm$0.12 & 22.92$\pm$0.18 & $<0.01$ & 1.19$\pm$0.11 & ... \\
J220431$+020140$ & 22:04:30.94 & $+$02:01:39.61 & 25.49$\pm$0.48 & 23.23$\pm$0.11 & 22.88$\pm$0.16 & 0.02 & 1.37$\pm$0.07 & ... \\
J220436$+015026$ & 22:04:36.49 & $+$01:50:26.46 & 25.24$\pm$0.38 & 23.20$\pm$0.10 & 22.14$\pm$0.10 & $<0.01$ & 1.32$\pm$0.04 & ... \\
J220748$+035644$ & 22:07:47.75 & $+$03:56:44.09 & 26.01$\pm$0.36$^{\rm b}$ & 23.37$\pm$0.13 & 22.42$\pm$0.07 & $<0.01$ & 1.19$\pm$0.11 & ... \\
J221034$+024506$ & 22:10:33.96 & $+$02:45:06.06 & 25.83$\pm$0.38$^{\rm b}$ & 23.17$\pm$0.09 & 22.28$\pm$0.11 & 0.33 & 1.40$\pm$0.07 & ... \\
J221529$+003846$ & 22:15:29.42 & $+$00:38:45.60 & 25.42$\pm$0.26 & 23.38$\pm$0.15 & 22.34$\pm$0.07 & $<0.01$ & 1.31$\pm$0.07 & ... \\
J221554$-005155$ & 22:15:54.37 & $-$00:51:55.22 & 25.57$\pm$0.33 & 23.47$\pm$0.10 & 22.79$\pm$0.11 & 0.01 & 1.31$\pm$0.13 & ... \\
J221725$-001220$ & 22:17:25.02 & $-$00:12:20.49 & 25.67$\pm$0.34 & 23.50$\pm$0.10 & 22.44$\pm$0.08 & $<0.01$ & 1.23$\pm$0.11 & ... \\
\enddata
\tablecomments{
Columns: (1) Candidate name. 
(2--3) Sky coordinates. 
(4--6) $i'$, $z'$, and $J$-band magnitudes with $1\sigma$ errors.  
(7) $w_{q}$ value determined from the photometric data from CFHTLS and IMS.  
(8) $i'_{\rm HSC}-z'_{\rm HSC}$ color in PSF magnitude from the HSC-SSP PDR3 catalog \citep{Aihara21}. 
(9) Spectroscopic identification.  if a quasar, the $z$ and $M_{1450}$ values are listed.
}
\tablenotemark{\tiny\rm{a}}{These are newly discovered quasars in this work.}

\tablenotemark{\tiny\rm{b}}{These are given in the $i'_2$ magnitudes.}
\end{deluxetable*}

\subsection{Initial Color Selection \label{sec:cs}}

The $\lya$ break of a $z\sim6$ quasar is located at $\lambda_{\rm obs}\sim8500$ \AA, giving a very red $i'-z'$ color with (almost) no detection at the shorter wavelengths.
On the other hand, at wavelengths longer than the $\lya$ emission, the quasar's color (e.g., $z'-J$) tends to be blue according to the quasar continuum emission.
Such colors are distinguished from those of late-type stars that are the main contaminants.
Lyman break galaxies (LBGs) can also be interlopers at $z\sim6$ \citep{Matsuoka16,Matsuoka18c}, but their expected number over our survey area is very small\footnote{The expected number density is calculated by integrating the LBG LF of \cite{Harikane21} down to $M_{1450}=-24$ ($-23.5$) mag, corresponding to $z'\sim23$ (23.5) mag. 
If we assume 100\% completeness, we obtain the expected number of 0.7 (4.8) by multiplying the cosmic volume of our survey area.
Like quasars, however, the survey completeness for LBGs is also expected to be very low at $-24<M_{1450}<-23.5$ mag. Therefore, we ignore them in the selection process.}
.
So, we ignore them in this study.

Figure \ref{fig:ccd} shows the color distributions of the point sources in our survey (gray contours).
Following \cite{Kim15}, we set the color and magnitude selection criteria as follows:

\begin{enumerate}
\item $i'-z'>2.0$
\item $z'-J<0.625\times((i'-z')+0.1)$
\item $u^{*}>u^{*}_{3\sigma}$, $g'>g'_{3\sigma}$, $r'>r'_{3\sigma}$
\item $z'<23.5$
\item $J<J_{5\sigma}$.
\end{enumerate}

\noindent 
Note that the magnitude with a subscript of $3\sigma$ ($5\sigma$) is the $3\sigma$ (5$\sigma$) limiting magnitude.
The first two color criteria are shown as the black solid lines in Figure \ref{fig:ccd}.
If a source is not detected at the $3\sigma$ level (e.g., $i'>i'_{3\sigma}$), then the limiting magnitude is used for the color selection instead.
Considering the point-source completeness (Section \ref{sec:pssel}) and the $i$-band limiting magnitude, we set the fourth criterion in terms of $z'$-band magnitude.
In addition, taking account of the variance in the $J$-band imaging depths (Figure \ref{fig:histdepth}), the $J$-band magnitude cut (the fifth criterion) is set at the 5$\sigma$ detection limit of the tiling image.

Among 404 color-selected objects, there are many spurious ones with bad image quality; most of them are located in the bad pixel regions in the image of at least one filter (e.g., at the edge of the image).
We automatically reject such cases, resulting in the 64 sources.
Then, we performed an additional visual inspection of the remaining sources to reject obvious noncelestial objects (e.g., diffraction spikes, bad pixels, image artifacts, cosmic rays, etc.)
We also cross-check the ones rejected by the above automated process and no object deserves to be selected through the visual inspection process.
We finally have 25 candidates, which are listed in Table \ref{tbl:cand}.

\subsection{AICc Selection \label{sec:aicc}}

It has been known that the observational properties of high-redshift quasars are slightly different from those of low-redshift ones.
For instance, the EW values of $z\sim6$ quasars tend to be smaller than those of low-redshift ones \citep{Banados16}.
Previous studies, however, used the low-redshift quasar templates which are redshifted to higher redshifts for statistical methods represented by the Bayesian approach to find high-redshift quasars (e.g., \citealt{Mortlock12,Matsuoka16}).
Concerned that this issue may miss plausible candidates, we here prefer to use the models whose parameters can be easily tuned to fit the observed properties.
Unlike when using observation-based templates, the complexity of the model and its potential for overfitting must be taken into account when using such models with multiple parameters.
We selected models that can represent the photometric characteristics of high-redshift quasars and late-type stars well with minimal parameters, which are described below.
Each model has a different number of free parameters, so we introduced an information criterion that prioritizes models for a given data set by giving an additional penalty based on the number of free parameters.
This approach is known to be effective in selecting the promising high-redshift quasar candidates by comparing models of different types of celestial objects \citep{Shin20}.
Moreover, we chose this method over the well-known Bayesian approach because it takes into account the ideal characteristics and distributions of the models, unlike observation-based templates.

In this study, we introduce the Akaike information criterion (AIC; \citealt{Akaike74}), which is based on the Kullback-Leibler discrepancy \citep{Kullback51}.
For a model $m$, AIC is given by

\begin{equation}
{\rm AIC}_{m} = 2k_{m} -2\ln(\mathcal{L}_{m}),\label{equ:aic}
\end{equation}

\noindent where $k_{m}$ is the number of free parameters and $\mathcal{L}_{m}$ is the likelihood.
The first term gives an additional penalty, allowing to compute the model priority with not only $\mathcal{L}_{m}$ but also $k_{m}$.
We have photometric information only in six bands, so a corrected version of AIC for small sample sizes (AICc; \citealt{Sugiura78}) works better than the equation (\ref{equ:aic}) \citep{Burnham02}, which is given by

\begin{equation}
{\rm AICc}_{m} = {\rm AIC}_{m} + \frac{2k_{m}(k_{m}+1)}{n-k_{m}-1}, \label{equ:aicc}
\end{equation}

\noindent where $n$ is the number of filters (or photometric data) to calculate $\mathcal{L}_{m}$.
By comparing the AICc values from different models, we can determine which model traces the observed data more closely.
Here, we introduce the two models: high-redshift quasars and late-type stars.

\subsubsection{High-redshift Quasar Model\label{sec:hzqmodel}}

We use the model of \cite{Kim19} based on the quasar composite spectrum of \cite{Vanden01}, including the IGM attenuation effect \citep{Madau96}.
It has four parameters of ($z$, $M_{1450}$, $\alpha_{p}$, EW), where $z$ is the redshift, $M_{1450}$ is the monochromatic magnitude at 1450 \AA, $\alpha_{p}$ is the slope of the quasar power-law continuum and EW is the equivalent width of the composition of $\lya$ and \ion{N}{5} emissions (see \citealt{Kim19} for details).
Instead of letting the parameters be free, we generated 0.1 million mock quasars that reflect the observational properties of real quasars at $z\sim6$.
The redshift and the magnitude are uniformly distributed (but randomly generated) in the ranges of $5.5<z<7.0$ and $-28<M_{1450}<-22$.
On the other hand, the other two parameters are randomly given by Gaussian distributions (mean$\pm$standard deviation); $\alpha_{P}=-1.6\pm1.0$ \citep{Mazzucchelli17} and $\log\rm{EW}=1.542\pm0.391$ \citep{Banados16}.
As in \cite{Kim20}, the Baldwin effect \citep{Baldwin77} is also included when we generate the EW distribution, by giving the shift to the mean value using the relation between EW of $\lya$ and continuum flux in \cite{Dietrich02}.
Although the relation is from low-redshift AGNs, we use it under the assumption of no redshift evolution in the quasar broad-line properties in rest-UV \citep{Shen19,Schindler20}.
From these model spectra, broadband magnitudes are calculated by integrating the mock quasar spectra convolved with the filter transmission curves.

\begin{figure*}
\centering
\epsscale{1.15}
\plotone{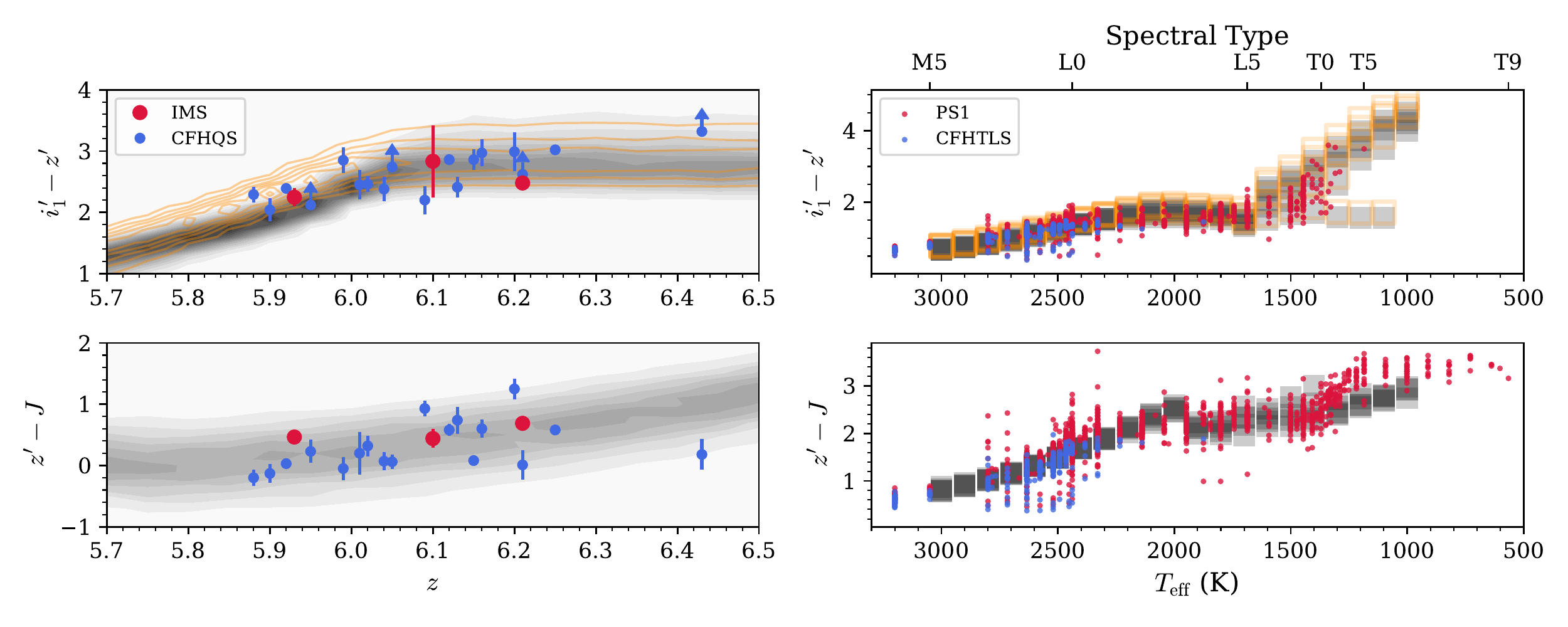}
\caption{
Left: Color distributions of the high-redshift quasar models across various redshifts (gray contours).
The over-plotted orange contour is for the case of $i'_2$-band magnitudes.
The red and blue circles represent the spec-identified quasars from IMS and CFHQS, respectively, while the arrows indicate the lower limit of colors.
Right: Color distributions of the late-type star model as a function of $T_{\rm eff}$ (gray squares).
The over-plotted orange open squares are for the case of $i'_2$-band magnitudes.
The red and blue circles represent the observed late-type stars in the PS1 and CFHTLS photometric systems, respectively (see details in Section \ref{sec:starmodel}).
\label{fig:colz}}
\end{figure*}

In the left panels of Figure \ref{fig:colz}, we show the color distributions of these mock quasars across the redshift.
As a function of redshift, they show good agreements with the confirmed quasars not only in this work (red circles) but also from the Canada-France-Hawaii Quasar Survey (CFHQS; \citealt{Willott07,Willott09,Willott10b}; blue circles).
This implies that the mock quasars emulate the real quasars at $z\sim6$ well.

\begin{figure*}
\centering
\epsscale{1.2}
\plotone{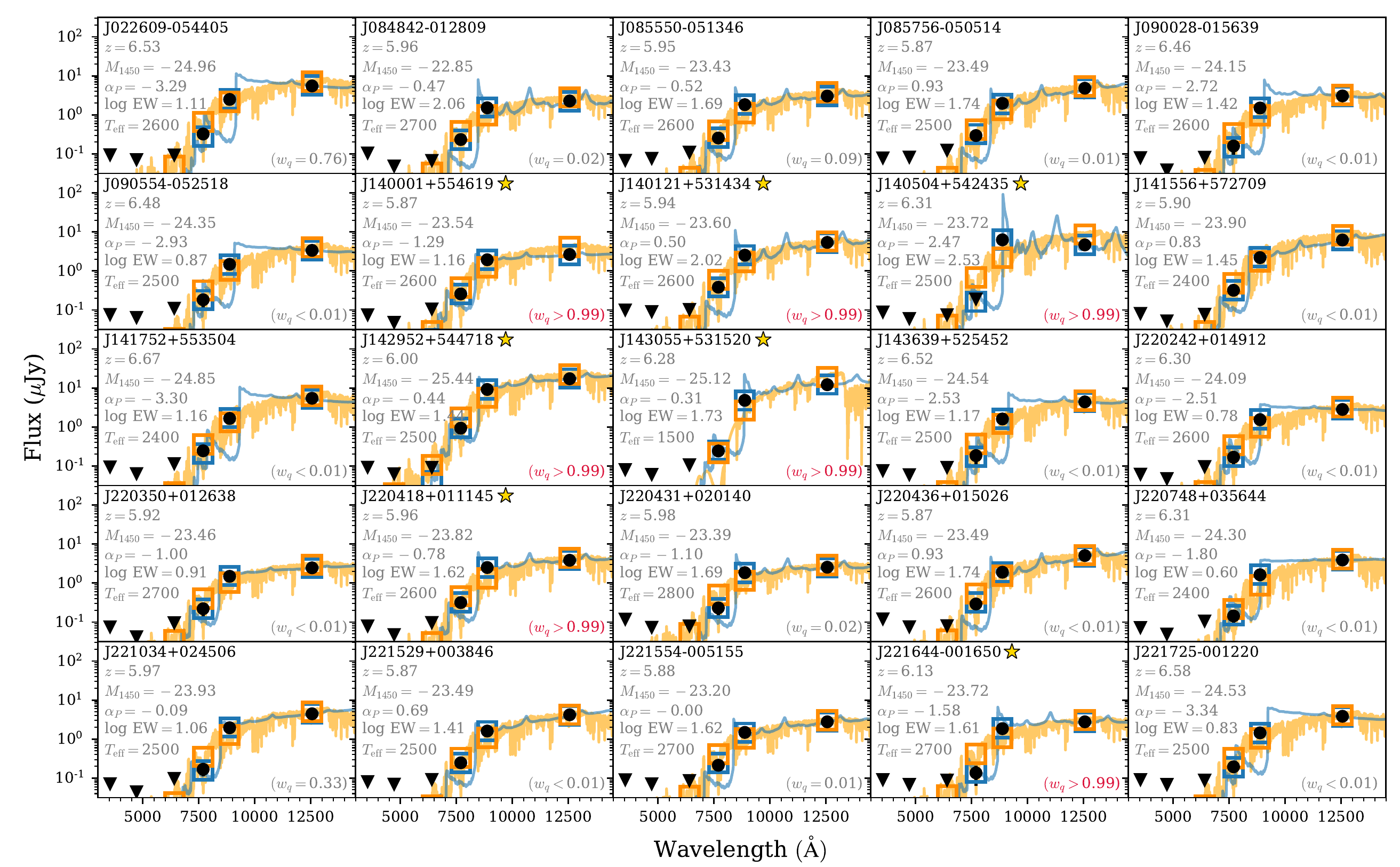}
\caption{
SED fitting results for the color-selected candidates.
The black circles and upside-down triangles are the photometric data points and their upper limits, respectively.
The blue and orange lines are the best-fit high-redshift quasar and late-type star models, respectively, while the open squares in the same colors represent fluxes in each band.
The key best-fit parameters of each candidate are marked in each panel.
The AICc-selected candidates with $w_{q}>0.99$ are highlighted with star symbols. 
\label{fig:sedfit}}
\end{figure*}

\subsubsection{Late-type Star Model\label{sec:starmodel}}

We use the BT-Settl model \citep{Allard12} for late-type stars, which is publicly available on the Theoretical Spectra Web Server\footnote{\url{http://svo2.cab.inta-csic.es/theory/newov2/index.php}}.
The model has four parameters: effective temperature ($T_{\rm eff}$), surface gravity ($g$), metallicity ([M/H]), and $\alpha$-element enhancement ($\alpha_{\rm E}$).
We choose the templates in the ranges of $1000~{\rm K}\leq T_{\rm eff} \leq 3000$ K and $3.5\leq\log(g)\leq5.5$ with step sizes of $\Delta T_{\rm eff}=100$ K and $\Delta\log(g)=0.5$, respectively.
Since the low $T_{\rm eff}$ ($\leq2500$ K) stars have a fixed value of [M/H]$=0$ and $\alpha_{\rm E}=0$, we only used the templates with those values.
Note that there is no template for a star with $T_{\rm eff}=1000$ K and $\log(g)=3.5$, resulting in the $30\times5+4=104$ templates.
In addition, we used a normalization factor $f_{N}$ as a free parameter. 
Like the quasar model, their magnitudes were obtained by integrating fluxes within each band.

The right panels of Figure \ref{fig:colz} show the color distribution of our late-type star model.
For comparison, we sourced the photometry of late-type stars from the Pan-STARRS1 3$\pi$ Survey (PS1) late-type star catalog \citep{Best18}.
The PS1 stellar spectral types were converted to $T_{\rm eff}$ using the relations between them \citep{Pecaut13,Bailey14}.
We also converted the PS1 magnitudes into the CFHTLS photometric system\footnote{\url{http://www.cadc-ccda.hia-iha.nrc-cnrc.gc.ca/en/megapipe/docs/filt.html}}, shown as the blue circles, except for the L- and T-dwarf stars without $g'_{\rm PS1}$-band photometry information.
The filter systems between the two surveys are only slightly different, so we also show the colors in PS1 as blue circles in order to see the trend of such L- and T-dwarf stars.
As in this figure, our late-type star models are broadly consistent with the real stars.

\subsubsection{SED Fitting and AICc-based Criterion \label{sec:sedfit}}

We performed the fitting for the spectral energy distributions (SEDs) of the 25 color-selected candidates in Section \ref{sec:cs} with the above high-redshift quasar and late-type star models.
As in \cite{Kim19}, for a model $m$, we find the best-fit solution that minimizes the modified $\chi^{2}$ statistic:

\begin{equation}
\chi^{2}_{m}=\sum_{d}\chi^{2}_{m,d}+\sum_{u}\chi^{2}_{m,u}.
\end{equation}

\noindent 
This statistic is to consider both detected and undetected cases (subscripts of $d$ and $u$, respectively).
The first term is a sum of typical $\chi^{2}$ statistic for the detected fluxes, given as

\begin{equation}
\chi^{2}_{m,d} = \left(\frac{F_{o,d}-F_{m,d}}{\sigma_{o,d}} \right)^{2},
\end{equation}

\noindent where $F_{o,d}$ is the observed flux in the $d$th band, $F_{m,d}$ is the model flux, and $\sigma_{o,d}$ is the uncertainty of $F_{o,d}$.
On the other hand, the second term gives an additional penalty for the cases of upper limit, defined by \cite{Sawicki12}:

\begin{equation}
\begin{aligned}
\chi^{2}_{m,u}& = -2 \ln \int_{-\infty}^{F_{{\rm lim},u}} \exp\left[ -\frac{1}{2} \left( \frac{F_{o,u}-F_{m,u}}{\sigma_{o,u}}\right)^{2}\right]dF\\
&=-2 \ln \left\{ \sqrt{\frac{\pi}{2}}\sigma_{o,u}\left[ 1 + {\rm erf}\left( \frac{F_{{\rm lim},u}-F_{m,u}}{\sqrt{2}\sigma_{o,u}} \right)\right] \right\},
\end{aligned} 
\end{equation}

\noindent where $F_{{\rm lim},u}$ is the upper limit of flux in the $u$th band, while $F_{o,u}$, $F_{m,u}$, and $\sigma_{o,u}$ are the observed flux, model flux, and the sensitivity in the same band, respectively.
${\rm erf}(x)$ is the error function of $(2/\sqrt{\pi})\int_{0}^{x}e^{-t^{2}}dt$, for the numerical calculation.

We calculate $\chi^{2}_{m}$ ($=-2\ln\mathcal{L}_{m}$) for the SEDs of the color-selected candidates.
For example, if a candidate is detected in the $i'$, $z'$, and $J$ bands, we calculate $\chi^{2}_{m,d}$ for these bands and $\chi^{2}_{m,u}$ for the other bands.
The best-fit quasar and star models, which minimize $\chi^{2}_{m}$ values, are shown as the blue and orange lines, respectively, in Figure \ref{fig:sedfit}.

Using Equations (\ref{equ:aic}) \& (\ref{equ:aicc}), we compute AICc$_{q}$ and AICc$_{s}$, where the subscripts $q$ and $s$ denote the high-redshift quasar and late-type star models, respectively.
Since our purpose is to determine whether a candidate is more likely to be a quasar or not, we only compare the best-fit cases from the two models.
To prioritize the models, we introduce the weights of AICc \citep{Burnham02}, given by 

\begin{equation}
w_{m}({\rm AICc})=\frac{l_{m}}{l_{q}+l_{s}},
\end{equation}

\noindent where

\begin{equation}
l_{m}=\exp\left(-\frac{1}{2}\left({\rm AICc}_{m}-{\rm AICc}_{\rm min}\right)\right)
\end{equation}

\noindent and ${\rm AICc}_{\rm min}$ is the minimum of the $\rm{AICc}$ values (min[AICc$_{q}$, AICc$_{s}$]).
This weight can be interpreted as the probability that the given model is the best one.
For the color-selected candidates, we listed their $w_q$ values in Table \ref{tbl:cand}.

We set a very strict criterion of $w_{q}>0.99$.
This corresponds to the fraction of the weights of $w_{q}/w_{s}\gtrsim100$, meaning that the high-redshift quasar model is $\gtrsim100$ times more likely to be the best model than the late-type star model \citep{Burnham02}.
This choice is because our late-type star model has a very small scatter in colors, compared to the observed ones, as shown in Figure \ref{fig:colz}.
There are seven candidates satisfying this criterion, shown as the orange filled circles in Figure \ref{fig:ccd}.

\subsection{Photometric Cross-check with HSC\label{sec:checkhsc}}

\begin{figure}
\centering
\epsscale{1.2}
\plotone{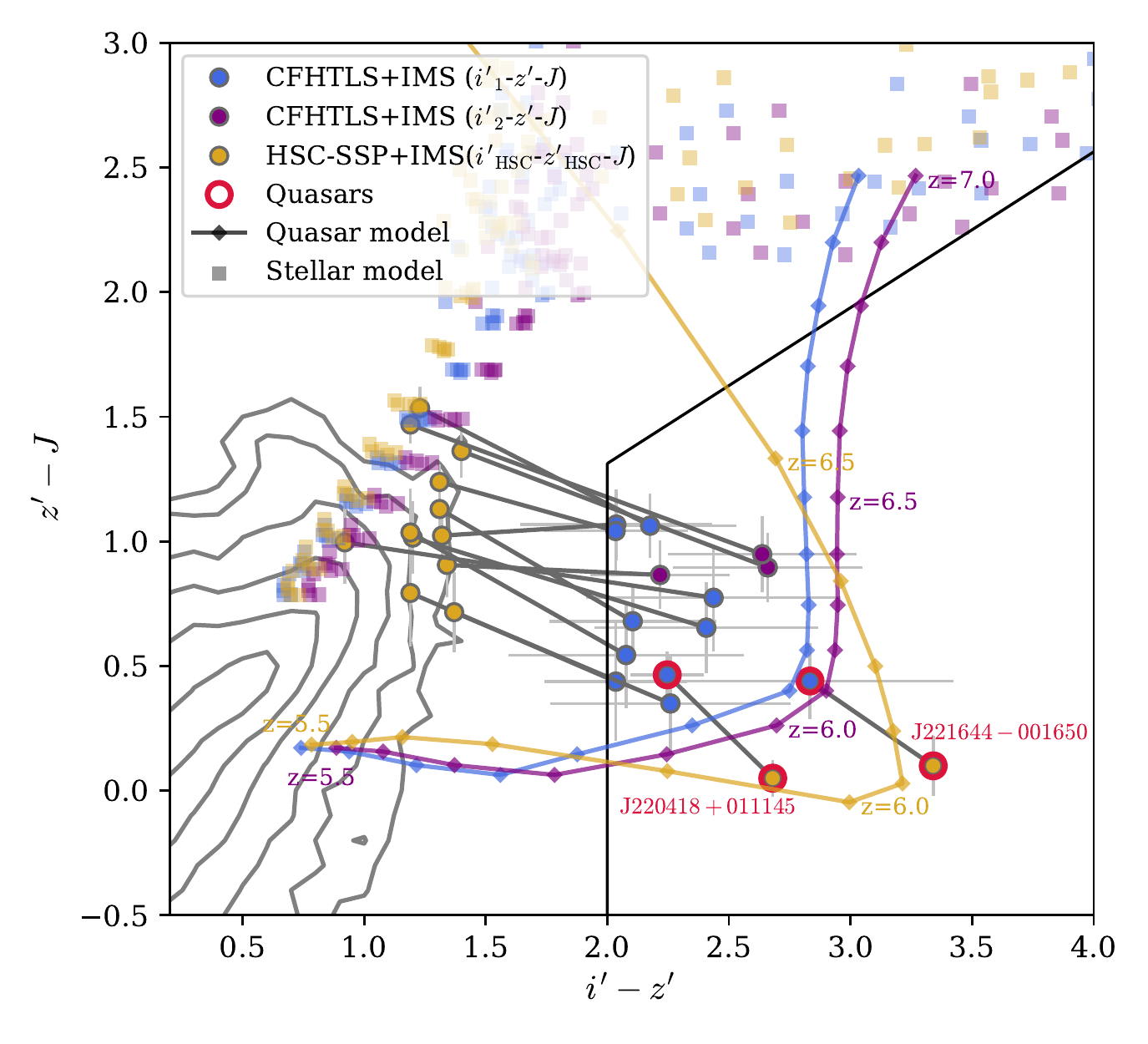}
\caption{
Color-color diagram of the 14 HSC-overlapped candidates (filled circles), similar to Figure \ref{fig:ccd}.
The blue, purple, and yellow colors represent the combinations of $i'_1$-$z'$-$J$, $i'_2$-$z'$-$J$, and $i'_{\rm HSC}$-$z'_{\rm HSC}$-$J$, respectively.
The diamonds with lines are the representative high-redshift quasar models, while the late-type star models are shown as squares with the given colors regardless of the temperature.
The gray lines indicate that the connected circles are the same candidates.
The spec-identified quasars are highlighted by red circles with their IDs.
\label{fig:hscizj}}
\end{figure}

Several parts of our survey area overlap with the area of the Wide Survey of HSC-SSP.
On average, the optical images from the HSC-SSP Public Data Release 3 (PDR3; \citealt{Aihara21}) are $\gtrsim 1$ mag deeper than those from CFHTLS.
Therefore, we expect more accurate photometry for the overlapping targets.

We found that 14 of the color-selected candidates are in the overlapping area, by matching our candidates to the sources in the HSC-SSP PDR3 catalog.
We adopted the HSC-SSP PDR3's PSF magnitudes in the $i'_{\rm HSC}$- and $z'_{\rm HSC}$-bands, listed in Table \ref{tbl:cand} in the $i'_{\rm HSC}-z'_{\rm HSC}$ form.
Note that the filter systems of CFHTLS and HSC are slightly different from each other, especially in the $z'$- and $z'_{\rm HSC}$ bands.
While their central wavelengths are similar to each other (8815 and 8908 $\rm\AA$, respectively), the former has a broader bandwidth of 1040 $\rm\AA$ than the latter, which has a bandwidth of 781 $\rm\AA$.
This makes different color trends of quasar on the color-color diagram, as shown in Figure \ref{fig:hscizj}; For our quasar model, the $i'_{\rm HSC}-z'_{\rm HSC}$ color (yellow diamonds) is redder at $z\sim6$ and becomes bluer at $z>6.5$ than $i'-z'$ (blue and purple diamonds).

Most of the HSC-overlapped candidates (12/14) have bluer colors of $i'_{\rm HSC}-z'_{\rm HSC}<1.5$. 
If we use $i'_{\rm HSC}$ and $z'_{\rm HSC}$ instead of $i'$ and $z'$, respectively, the 12 candidates move toward the stellar locus in the color-color diagram (yellow circles), meaning that they are unlikely to be $z\sim6$ quasars or even galaxies at that redshift \citep{Harikane21}.
It is remarkable that all of them are also rejected by our AICc selection using the CFHTLS optical photometry.
On the other hand, the colors of the other two candidates, highlighted by red circles, are still likely to be those of the high-redshift quasar models even if using the HSC colors.
We note that they satisfy our AICc-selection criterion and were also identified as high-redshift quasars by spectroscopy \citep{Kim15,Matsuoka16}.
This demonstrates that our AICc selection, even under shallower imaging data, is an effective method to exclude nonquasar objects.

On the contrary, there may be objects that have red $i'_{\rm HSC}-z'_{\rm HSC}$ colors in HSC but not so in our data.
From the HSC-SSP PDR3 catalog, we select point sources with $i'_{\rm HSC}-z'_{\rm HSC}>2$ using the selection criteria given in equation (1) of \cite{Matsuoka18c}, which are also detected in CFHTLS $z'$-band images.
There are seven isolated point-sources at $z'<23.5$ mag, while the brightest one among them has $i'-z'=1.58$ in our data.
This is because of its brighter $i'$-band magnitude in IMS ($i'=24.51\pm0.18$ vs $i'_{\rm HSC}=25.00\pm0.10$), while there is no significant difference in $z'$-band magnitudes ($z'=22.93\pm0.05$ mag vs $z'_{\rm HSC}=22.88\pm0.03$).
We note that this object is close to the edge of the CFHTLS image.
This implies that 14\% (1/7) of red objects could be missed in our imaging data especially for $z'<23.5$ mag objects.

\section{SPECTROSCOPIC IDENTIFICATION\label{sec:identify}}

In previous studies, three of our candidates were already identified as $z\sim6$ quasars: J142952$+$544718 \citep{Willott10b}, J220418$+$011145 \citep{Kim15,Kim18}, and J221644$-$001650 \citep{Matsuoka16}.
For the remaining targets, we additionally obtained their spectra with the Palomar 200 inch and the Gemini telescopes.

\begin{figure*}
\centering
\epsscale{1.2}
\plotone{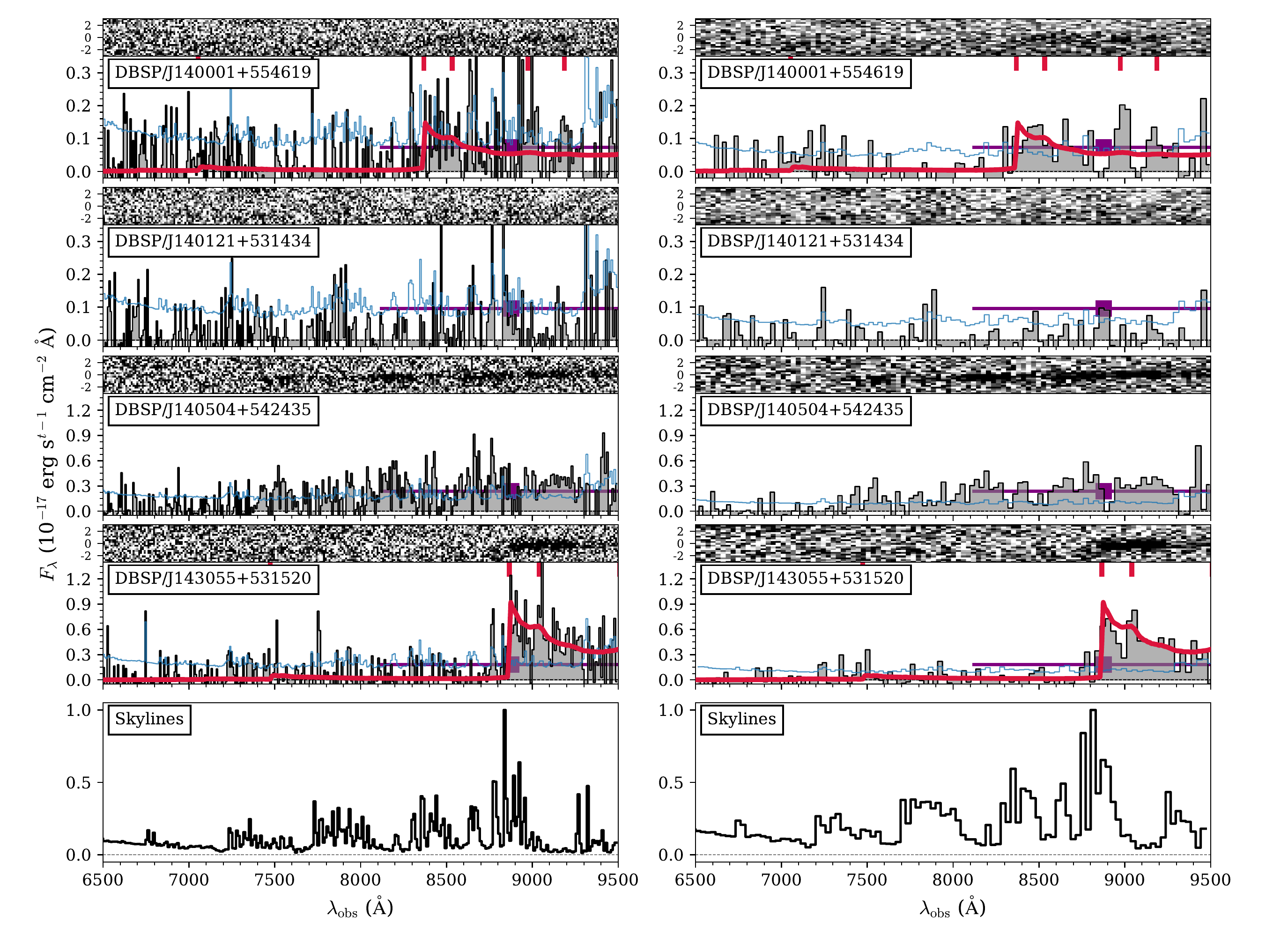}
\caption{
P200 optical spectra of four quasar candidates, after binning with resolutions of $R=956$ (instrumental setup; left) and 300 (right).
The blue lines represent the flux uncertainties in 1$\sigma$ level.
The red lines on the spectra of J140001$+$554619 and J143055$+$531520 are their best-fit high-redshift quasar models of which emission line locations are marked as the red vertical ticks (Ly$\beta$ $\lambda1025$, $\lya$ $\lambda1216$, \ion{N}{5} $\lambda1240$, \ion{O}{1} $\lambda1304$, and \ion{C}{2} $\lambda1335$ from left to right).
The purple squares show the $z'$-band fluxes with its bandwidth.
The bottom panels show the normalized skylines.
Note that the y-axis of the 2D spectra is given in units of arcsec.
\label{fig:spec_p200}}
\end{figure*}

\begin{figure}
\centering
\epsscale{1.2}
\plotone{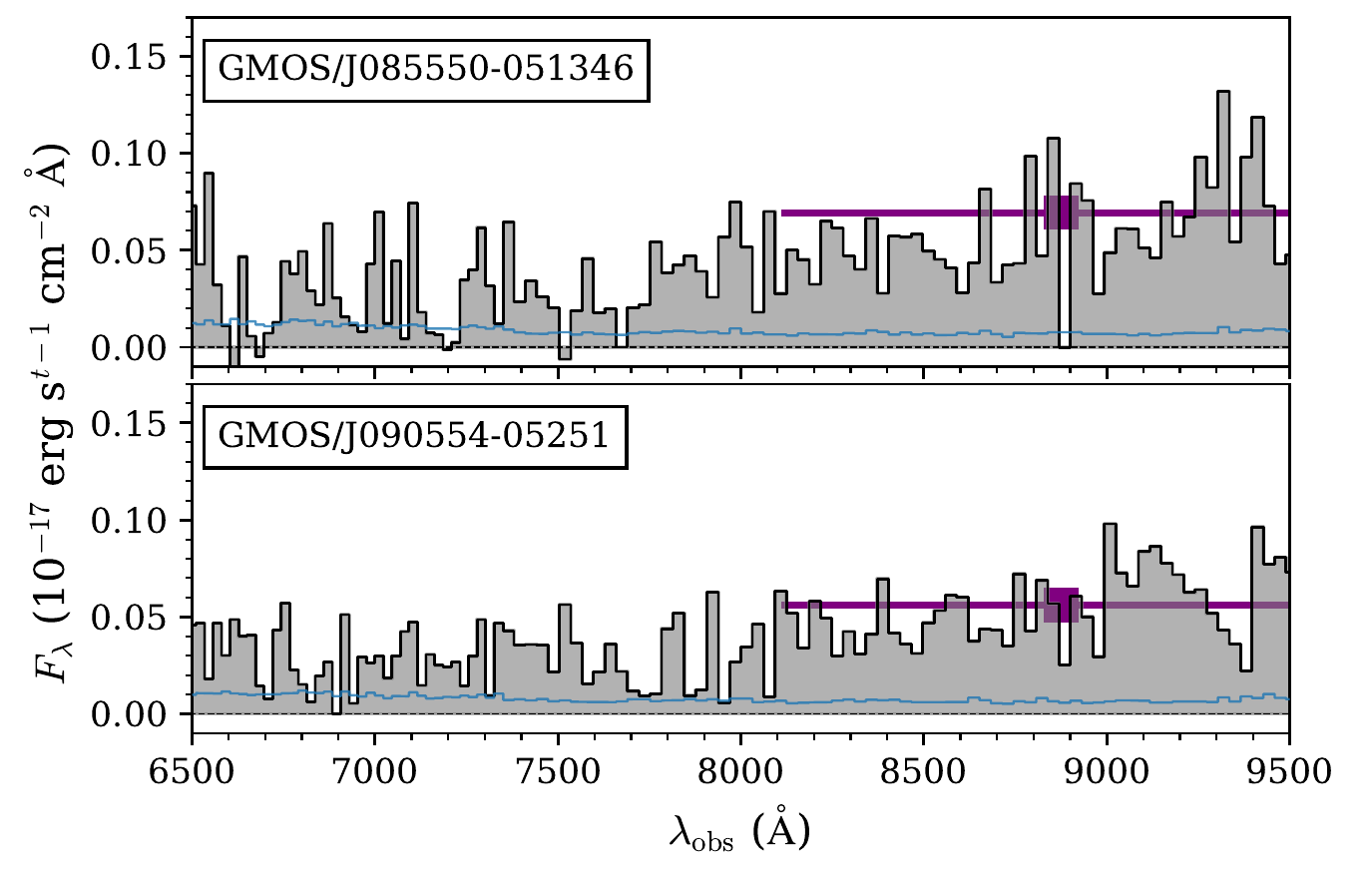}
\caption{
GMOS optical spectra of two quasar candidates, dissatisfying the AICc criterion. The symbols are same as in Figure \ref{fig:spec_p200}.
\label{fig:spec_gmos}}
\end{figure}

\subsection{P200/DBSP Observation}

We carried out spectroscopic observations of the other four AICc-selected candidates with the Double Spectrograph (DBSP) on the Palomar Hale 200 inch telescope (P200) on 2021 July 13 (PID:CTAP2021-A0032), under the seeing condition of $\sim1\farcs5$.
We used the grating of 316 lines/mm with a $1\farcs5$-width slit, giving the resolution of $R=956$.
To avoid the duplication of the 0th order spectrum, the D55 dichroic filter ($\lambda_{\rm obs}>5500$ \AA) was used.
The total exposure times are 3600 s for fainter ones (J140001$+$554619 \& J140121$+$531434) and 1200 s for the other brighter ones (J140504$+$542435 \& J143055$+$531520). 

For data reduction, we used the PypeIt Python package\footnote{\url{https://pypeit.readthedocs.io/}} \citep{Prochaska20a,Prochaska20b}.
This is an open-source pipeline for the selected instruments, which automatically performs the bias subtraction, flat fielding, sky-line subtractions, and wavelength calibrations (with HeNeAr arc lines).
Considering the faintness of our target, we manually extracted fluxes within an optimal aperture with a fixed FWHM that matches the seeing size ($1\farcs5$).
The flux calibration was also done by PypeIt with the standard star, Feige110.
In addition, we scaled the spectra to match with their $z'$-band photometry to compensate for the flux loss by sky fluctuation, as was done in \cite{Kim19} and \cite{Kim20}.
By convolving the $z'$-band transmission curve with the three spectra (except for J140121$+$531434 without detection), we obtained scaling factors of 1.54--1.78.
We applied the average scaling factor of 1.69 to all the DBSP spectra.
Note that the limited wavelength coverage of our spectra up to $\sim10,000$ \AA~may affect the scaling factor.
Finally, we binned the spectra in the spectral direction with resolutions of $R=956$ and 300 by using the inverse-variance weighting method (e.g., \citealt{Kim18}).

We show the DBSP spectra of the four candidates in Figure \ref{fig:spec_p200}.
Except for the faintest J140121$+$531434, their spectra are marginally detected with low S/N of 2--3.
J140001$+$554619 and J143055$+$531520 show clear breaks at $\sim8400$ and $8850$ \AA, respectively.
Such breaks are more clearly visible if we maximize the S/N by binning the data to a low resolution of $R=300$ (right columns in Figure \ref{fig:spec_p200}).
We provide more detailed individual notes for these targets in Section \ref{sec:specfit}.

As mentioned above, J140121$+$531434 is not detected, even though its $z'$-band magnitude is brighter than J140001$+$554619.
For a high-redshift quasar, the peak of $\lya$ flux in its spectrum is expected to be brighter than broadband photometry ($z'$-band; purple squares) which is dominantly determined by continuum emission, but this object shows no remarkable feature.
On the other hand, the spectrum of J140504$+$542435 has continuum emissions without any remarkable emission lines or breaks.
Therefore, we concluded that J140121$+$531434 and J140504$+$542435 are not high-redshift quasars, but interlopers like late-type stars or faint galaxies.

\subsection{Gemini/GMOS Observation}

We obtained the spectra of J085550$-051346$ and J090554$-$052518 with the Gemini Multi-Object Spectrograph (GMOS; \citealt{Hook04}) on the Gemini-South 8 m Telescope on 2020 February 24 (PID: GS-2020A-Q-219).
Note that these observations preceded the AICc selection.
The seeing condition was $\sim0\farcs7$.
Since the targets are very faint, we optimized the observing configurations to maximize S/N.
The choice of an R150\_G5326 grating with a slit width of $1\farcs0$ gives a low resolution of $R\sim315$, and we set the $4\times4$ binning in the spatial/spectral directions.
The nod-and-shuffle mode was used to subtract skylines accurately.
The total exposure times are 2904 and 3388 s for J085550$-051346$ and J090554$-$052518, respectively.

We followed the general reduction process for the GMOS spectra using the Gemini IRAF package: (1) bias subtraction, (2) flat-fielding, (3) sky-line subtraction, (4) wavelength calibration with CuAr arc lines, and (5) flux calibration with a standard star LTT2415.
Note that for the extraction process, we used a fixed aperture whose size is consistent with the seeing size of $\sim0\farcs7$.
Like DBSP spectra, the GMOS spectra were scaled to match with $z'$-band magnitudes, by a factor of $2.25$ on average.
Such a large scaling factor might be due to the reported problem on the coefficient of thermal expansion during our observing run\footnote{\url{https://www.gemini.edu/news/instrument-announcements/gmos-s-data-affected-ccd1-cte-problem}}.
We also binned the spectra along the spectral direction to match the instrument resolution of $R=315$.

The two spectra are shown in Figure \ref{fig:spec_gmos}.
Both show clear continua but without any remarkable emission lines or $\lya$ break, suggesting that they are not high-redshift quasars.
We here emphasize that they dissatisfy the AICc criterion, supporting the feasibility of our approach.

\begin{deluxetable*}{lccccccccc}
\tabletypesize{\scriptsize}
\tablecaption{Best-fit Parameters of IMS $z\sim6$ Quasars\label{tbl:sample}}
\tablehead{
\colhead{ID} & \multicolumn{4}{c}{Spectroscopy} &  \multicolumn{4}{c}{Photometry} & \colhead{Spectral Reference}\\
\cmidrule(lr){2-5} \cmidrule(lr){6-9}
 & \colhead{$z$} & \colhead{$M_{1450}$} & \colhead{$\alpha_{P}$} & \colhead{$\log{\rm EW}$} &  \colhead{$z$} & \colhead{$M_{1450}$} & \colhead{$\alpha_{P}$} & \colhead{$\log{\rm EW}$} & 
}
\startdata
J140001$+554619$ & 5.85 & $-23.25$ & $-1.49$ & $1.54$ & 5.87 & ${-23.54}$ & $-1.29$ & 1.16 &This work\\
J143055$+531520$ & 6.29 & $-25.45$ & $-0.80$ & $1.56$ & 6.28 & ${-25.12}$ & $-0.31$ & 1.73 & This work\\
\hline
J142952$+544718$ & {6.21} & ${-25.85}$ & ... & ... & 6.00 & ${-25.44}$ & $-0.44$ & 1.44 & \cite{Willott10b}\\
J220418$+011145$ & {5.93} & ${-23.99}$ & ... & ... & 5.96 & ${-23.82}$ & $-0.78$ & 1.62 & \cite{Kim15,Kim18}\\
J221644$-001650$ & {6.10} & ${-23.56}$ & ... & ... & 6.13 & ${-23.72}$ & $-1.58$ & 1.61 & \cite{Matsuoka16}\\
\enddata
\tablecomments{This table provides the parameters of the high-redshift quasar models (Section \ref{sec:hzqmodel}), which are best-fitted for the spectroscopy and photometry of the IMS $z\sim6$ quasars, respectively.}
\end{deluxetable*}

\subsection{Spectral Properties of New Quasars\label{sec:specfit}}

For J140001$+$554619 and J143055$+$531520, which are likely to be high-redshift quasars with clear breaks on their spectra, we measured their $z$ and $M_{1450}$ by finding the best-fit models among our mock quasars in Section \ref{sec:hzqmodel}.
We calculated the reduced chi-square values ($\chired$) between their spectra and our mock quasar spectra.
The wavelength range for the spectral fitting was set as 6500 \AA~ $<\lambda_{\rm{obs}}<$ 9500\,\AA.
We chose the best-fit models of J140001$+$554619 and J143055$+$531520 with the minimum $\chired$ values of $\chi^{2}_{\rm red,min}=1.01$ and 1.09, respectively, which are shown as the red lines in Figure \ref{fig:spec_p200}.
The best-fit parameters are listed in Table \ref{tbl:sample}, along with the results from the photometric data (Section \ref{sec:sedfit}).
We also list the measurements of the three previously known quasars at $z\sim6$ \citep{Willott10b,Kim15,Kim18,Matsuoka16}.

The clear break of J140001$+$554619 at $\sim8400$ \AA ~is consistent with the $z=5.85$ quasar model.
Interestingly, some peaky detections on the spectrum are also in line with the locations of quasar's high-ionization emission lines: \ion{O}{1} $\lambda 1304$, and \ion{C}{2} $\lambda 1335$ (red vertical markers in Figure \ref{fig:spec_p200}).
The existence of the probable emission lines may provide additional evidence supporting its nature as a high-redshift quasar.
However, since the S/N of the two peaky detections are as low as 2.7 and 1.9, respectively, we still have doubts about their reliability.
LBGs would have a clear Lyman break on their spectra too, so it is difficult to confidently say that J140001$+$554619 is not a $z\sim6$ LBG, although we have ignored them in our selection process.
Further observations are needed to identify the high-ionizing emission lines, which can be a crucial criterion for determining whether this object is a quasar or LBG.
In the following sections, we assume that J140001$+$554619 is a high-redshift quasar, but we here caution that our estimates could be overestimated if this object is actually an LBG.

In the case of J143055$+$531520, there is a clear $\lya$ break with a plausible \ion{N}{5} $\lambda 1240$ emission line (S/N$\sim7$. calculated from peaky emissions at $9010-9072$ $\rm\AA$ on the $R=300$ spectrum), which is consistent with the $z=6.29$ quasar model.
With little chance of finding LBGs as bright as this target ($M_{1450}=-25.12$ mag; inferred from the LBG LF of \citealt{Harikane21}), we conclude that J143055$+$531520 is a high-redshift quasar.

The photometric redshifts ($z_{\rm phot}$) of the five spectroscopically identified quasars are well in line with the spectroscopic redshifts ($z_{\rm spec}$); the standard deviation of $\delta z=(z_{\rm phot}-z_{\rm spec})/(1+z_{\rm spec})$ is only 0.013.
This is much smaller than the value for $z\sim5$ quasars using the same model (0.043; \citealt{Kim19}), which might be due to the stronger IGM attenuation at higher redshifts, despite the small number statistics.
The DBSP spectra with low S/N of 2--3 and limited wavelength ranges of $\lambda_{\rm obs}<9500$\,\AA~gives a degeneracy of $M_{1450}$, $\alpha_{p}$, and EW in the fitting at $\lambda_{\rm obs}\gtrsim 9000$\,\AA.
Meanwhile, our SED fitting for photometry includes $z'$- and $J$-band magnitudes, enabling us to estimate $M_{1450}$ better.
Therefore, we use their magnitudes from photometry instead of those from spectra in the following analysis.

\section{QUASAR SPACE DENSITY \label{sec:qlf}}

\subsection{Survey Completeness\label{sec:comp}}

As mentioned in Section \ref{sec:overlap}, the imaging depths of our survey are not uniform (Figure \ref{fig:histdepth}).
Therefore, we calculated the completeness for every 1 arcmin$^{2}$ area (herafter ``patch''), given as a function of $z$ and $M_{1450}$: $f_{X,p}(z,M_{1450})$, where $X$ is the type of completeness and $p$ is the index of each patch.
We used bin sizes of $dz=0.05$ and $dM_{1450}=0.2$ mag, including more than one hundred mock quasars (Section \ref{sec:hzqmodel}) in each bin.

\subsubsection{Detection and Point-source-selection Completeness \label{sec:comp1}}

\begin{figure}
\centering
\epsscale{1.2}
\plotone{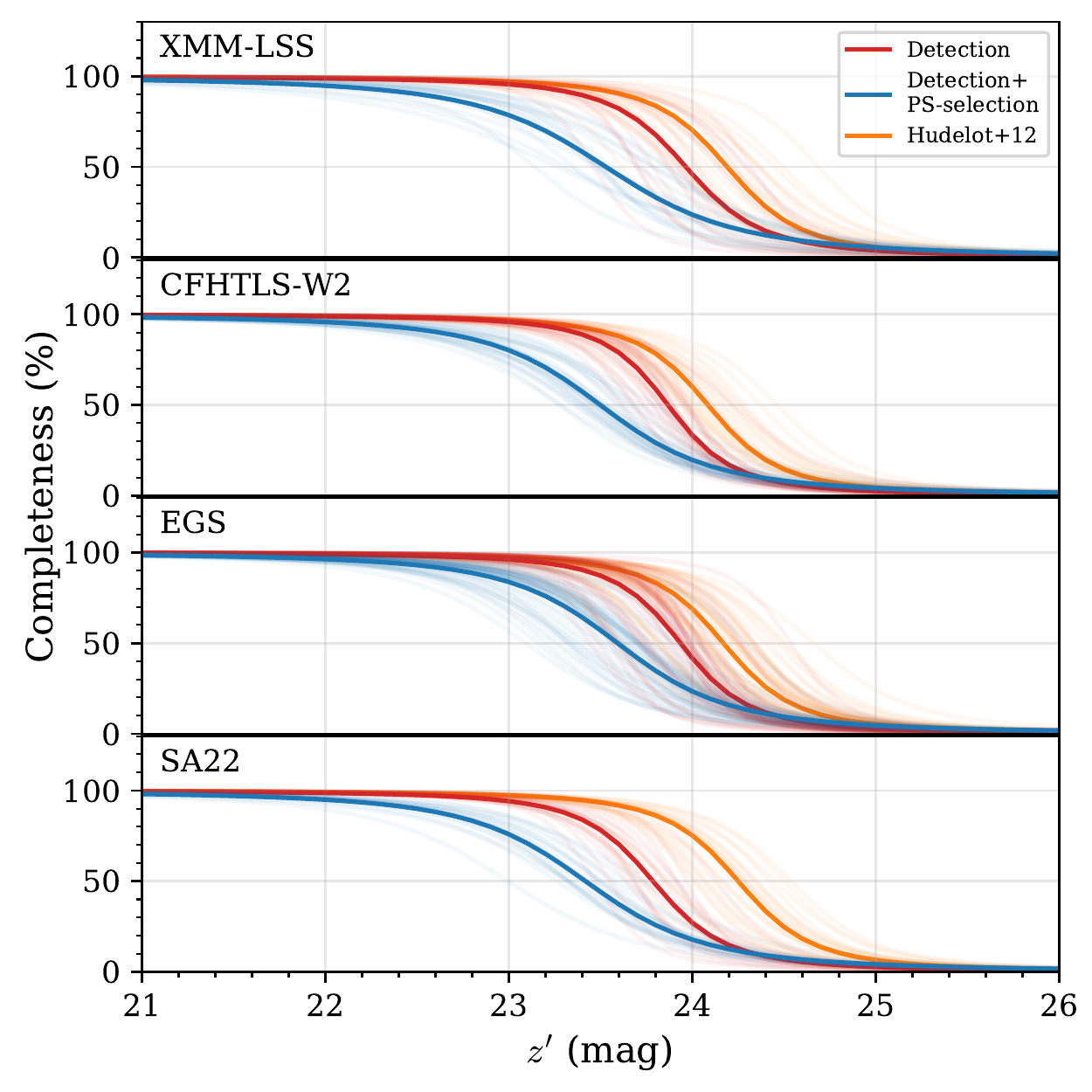}
\caption{
Point-source detection completeness of our $z'$-band images in the four survey fields.
The red and blue lines are the results of including our detection criteria and detection+point-source selection criteria, respectively.
The orange lines are the completeness function by \cite{Hudelot12} that used more lenient detection criteria.
The translucent lines denote the completeness of each CFHTLS tile, while the median values are highlighted by the thick solid lines.
\label{fig:photcomp}}
\end{figure}

We first consider the photometric completeness related to the source detection.
We used the artificial stars from a simulation described in Section \ref{sec:pssel} to test how many of them can be recovered with our images and methods along with the magnitude (Section \ref{sec:extract}).
The resultant completeness is parameterized by the equation of \cite{Fleming95}:

\begin{equation}
c(z')= \frac{1}{2}\left( 1 - \frac{\alpha_{50}(z'-z'_{50})}{\sqrt{1+\left(\alpha_{50}(z'-z'_{50})\right)^2}} \right)\label{equ:fleming},
\end{equation}

\noindent where $\alpha_{50}$ is the slope of the function at the 50\% completeness magnitude ($z'_{50}$).
The results for each CFHTLS tile are shown as the red lines in Figure \ref{fig:photcomp}, while the median values are highlighted by the thick solid line.

For comparison, we also parameterized the photometric completeness of \cite{Hudelot12} with Equation (\ref{equ:fleming}), shown as the orange lines.
Note that the functions were shifted in the magnitude direction by following our new $zp$ measurements.
Our result show lower $z'_{50}$ ($\lesssim24$ mag) than \cite{Hudelot12}.
This is because of our choice of SExtractor parameters for searching high S/N sources: $\texttt{DETECT\_MINAREA}=9$ and $\texttt{DETECT\_THRESH}=1.3$.
These are more stringent than $\texttt{DETECT\_MINAREA}=3$ and $\texttt{DETECT\_THRESH}=1.0$ used by \cite{Hudelot12}.
If we use these values instead, our simulation gives consistent results with \cite{Hudelot12}.
But we point out that there is only a negligible difference in detection rate ($<1\%$) between ours and \cite{Hudelot12} at $z'<23.5$ mag.

In addition to the detection completeness, we consider the point-source selection described in Section \ref{sec:pssel}.
Using Equation (\ref{equ:fleming}), we also fitted the binned completeness for the detected sources satisfying our point-source selection criterion.
The results are shown as the blue lines in Figure \ref{fig:photcomp}, which have $z'_{50}\sim23.5$ mag on average, which is naturally lower than those of the detection completeness limits.
This means that our magnitude cut of $z'<23.5$ mag is very marginal.
Using the mock quasar sample described in Section \ref{sec:hzqmodel}, we converted the completeness for a given patch to a function of $z$ and $M_{1450}$: $f_{{\rm D},p}(z,M_{1450})$.

\subsubsection{Color-selection Completeness\label{sec:selfun}}

\begin{figure*}
\centering
\epsscale{1.15}
\plotone{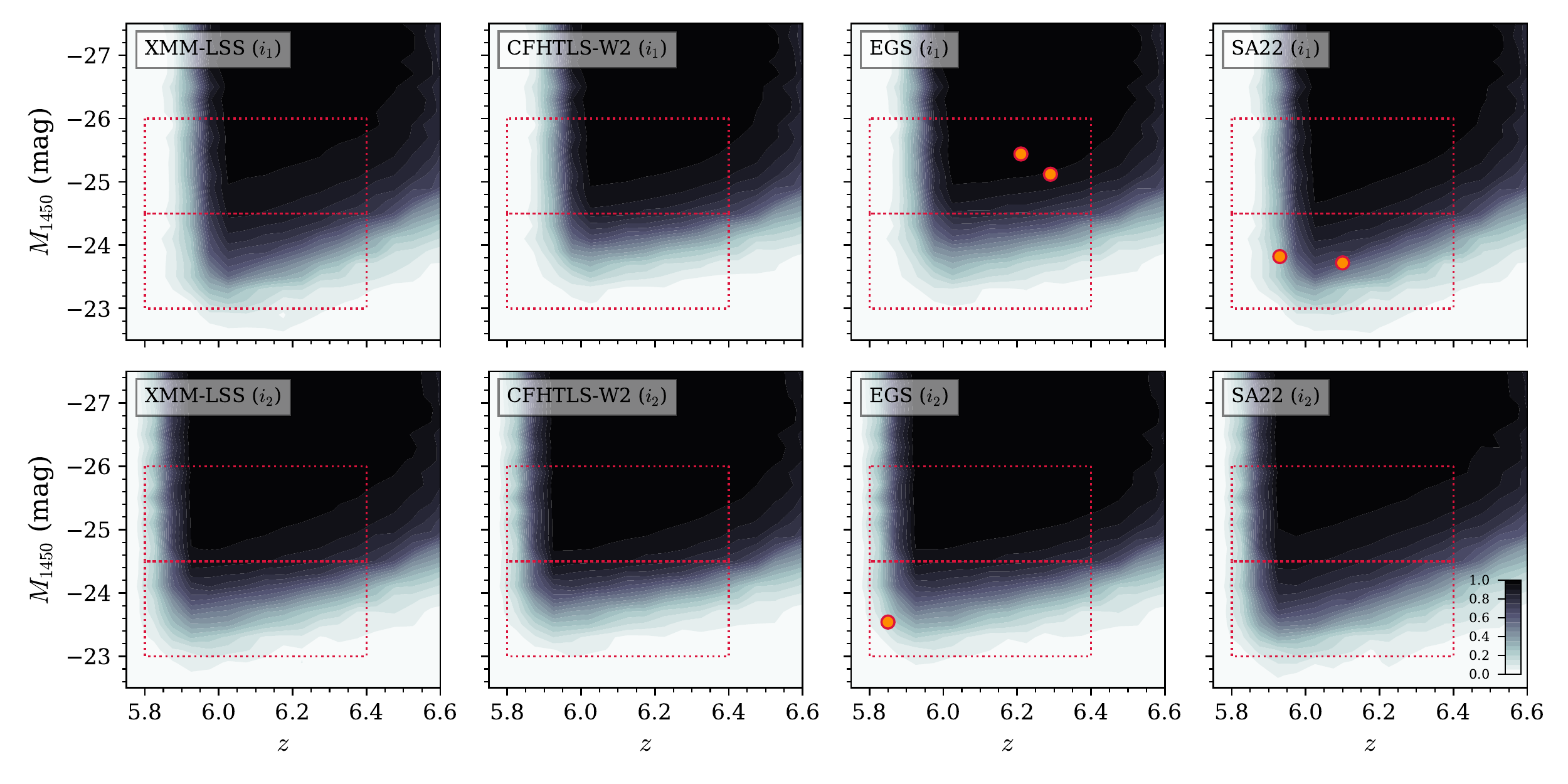}
\caption{
Selection completeness in each field as a function of $z$ and $M_{1450}$, $F_{\rm field} (z,M_{1450})$, divided into the cases of $i'_{1}$ (top row) and $i'_{2}$ (bottom row) bands.
The colorbar shows the completeness level.
The orange circles are the spec-identified quasars, listed in Table \ref{tbl:sample}.
The red boxes indicate the two bins for estimating space density.
\label{fig:selfun}}
\end{figure*}

\begin{figure}
\centering
\epsscale{1.15}
\plotone{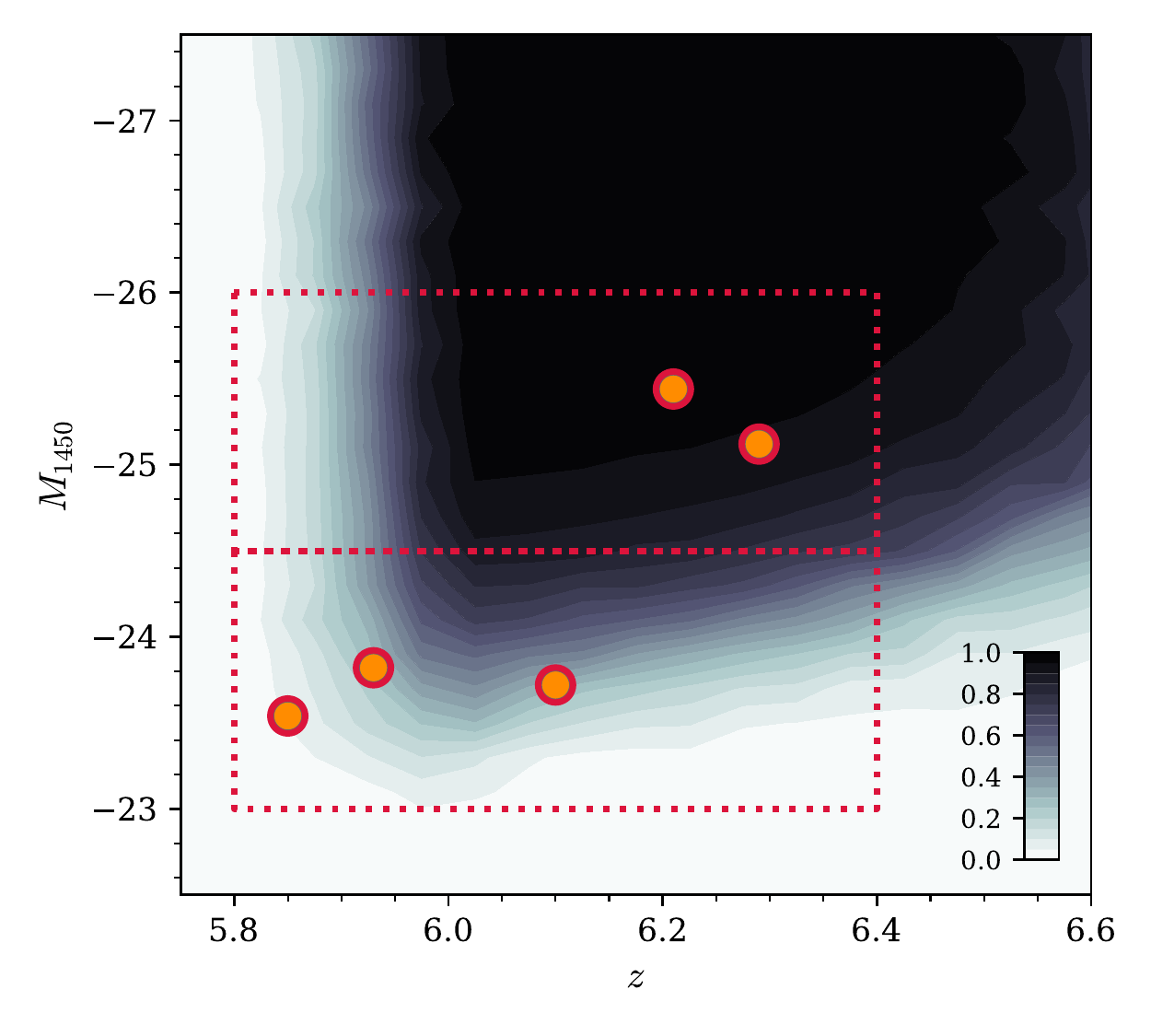}
\caption{
Total selection completeness as a function of $z$ and $M_{1450}$.
The symbols are same as in Figure \ref{fig:selfun}.
The red boxes indicate the two bins for estimating space density.
\label{fig:selfun_comb}}
\end{figure}

Our initial selection is based on the colors, so we calculate the quasar selection efficiency of our color-selection criteria described in Section \ref{sec:cs}.
We gave random Gaussian noises to the magnitudes of the mock quasars according to the imaging depths at a given patch.
Then, the fraction of quasars satisfying the criteria in each $(z,M_{1450})$ bin was calculated, resulting in the color-selection completeness of $f_{{\rm C},p}(z,M_{1450})$.
Note that the difference between the $i'_1$- and $i'_2$-band images are also considered.

\subsubsection{AICc-selection Completeness \label{sec:aiccselfun}}

We considered the application of the AICc selection for the final candidates.
The fraction of the mock quasars satisfying $w_{q}>0.99$ was calculated patch by patch as in Section \ref{sec:selfun}.
Since the mock quasars have no error information in their magnitudes, we gave appropriate magnitude errors according to their magnitudes and imaging depths in each patch.
The resultant completeness,  $f_{{\rm A},p}(z,M_{1450})$, shows 90\% down to $M_{1450}\sim-23.5$ mag, meaning that the AICc selection do not reduce the total selection completeness significantly at the magnitude ranges of interest.

\subsubsection{Total Selection Completeness \label{sec:totalselfun}}

At a given patch, the total selection completeness is calculated by multiplying the above completeness functions because they are independent with each other: $f_{p}=f_{{\rm D},p}\times f_{{\rm C},p}\times f_{{\rm A},p}$. 
Then we combined $f_{p}$ in each field to get the average completeness; $F_{\rm field}(z,M_{1450}) =\left[\sum_{\rm field}f_{p}(z,M_{1450})\right]/N_{\rm field}$, where $N_{\rm field}$ is the total number of the patches in the field.
Figure \ref{fig:selfun} shows the resultant completeness $F_{\rm field}(z,M_{1450})$ of each field.
As can be inferred from the quasar track in Figure \ref{fig:ccd}, the difference between the $i'_{1}$ and $i'_{2}$ band filters are reflected in the results; the usage of the $i'_{2}$ band filter can catch more quasars at $z<5.9$.
The two brighter quasars are in the parameter space where the completeness is $\sim1$, while the remaining three fainter ones have completeness values of 0.1--0.3.
But the three fainter quasars are in the low completeness region is due to them being found in the survey area of the SA22 and EGS fields where deeper $J$-band images are available.
Indeed, $f_{p}(z,M_{1450})$ for the three quasars is 0.2--0.5, which rather deserves to be selected.
We also calculated the total completeness ($F(z,M_{1450})=\left[ \sum f_{p}(z,M_{1450}) \right]/N_{p}$), shown in Figure \ref{fig:selfun_comb}.

\subsection{Binned Space Density \label{sec:binqlf}}

As listed in Table \ref{tbl:sample}, we have five $z\sim6$ quasars identified by spectroscopy within the IMS survey area, including the two new quasars in this work.
Their $z$-$M_{1450}$ distributions are shown as the orange filled circles in Figure \ref{fig:selfun}.
With this complete sample of $z\sim6$ quasars,
we calculate the binned space density using the $1/V_{a}$ method of \cite{Avni80}, where $V_{a}$ is the specific comoving volume.
For given bin sizes of $\Delta M_{1450}$ and $\Delta z$, $V_{a}$ can be calculated as

\begin{equation}
V_a = \frac{1}{\Delta M_{1450}} \iint F(z,M_{1450})\, \frac{dV_c}{dz}\,dz\,dM_{1450},\label{equ:va}
\end{equation}

\noindent where $dV_{c}/dz$ is the comoving element of our survey area.
Then we calculate the binned space density ($\Phibin$) and its error ($\sigma_{\Phibin}$) as following:

\begin{equation}
\Phibin (M_{1450}) = \frac{1}{\Delta M_{1450}} \sum^{N_{\rm bin}}\,\frac{1}{V_a}\label{equ:phibin},
\end{equation}

\noindent and

\begin{equation}
\sigma_{\Phibin} (M_{1450}) = \frac{1}{\Delta M_{1450}} \left[ \sum^{N_{\rm bin}}\,\left(\frac{1}{V_a}\right)^{2}\right]^{1/2},
\end{equation}

\noindent where $N_{\rm bin}$ is the number of objects in the given bin.
This method critically depends on the choice of the bin.
Considering the small number of our sample, we set a single redshift bin of $5.8<z<6.4$.
The average redshift of our sample is $z=6.08$.
Meanwhile, we took two large $M_{1450}$ bins: $-26.0\leq M_{1450}<-24.5$ and $-24.5\leq M_{1450}<-23.0$ (red boxes in Figure \ref{fig:selfun} \& \ref{fig:selfun_comb}).
Such large bins in $M_{1450}$ were chosen because our sample is small but complete.
There are two and three quasars in each bin, and their average magnitudes are $M_{1450}=-25.28$ and $-23.69$ mag, respectively.
The resultant $\Phibin$ values are listed in Table \ref{tbl:binqlf} (top two rows).
We note that the average redshifts of the two bins are $z=6.25$ and 5.97, respectively.
The discrepant redshift values reflect the shape of the completeness function and the larger volume available for higher redshifts for the brighter sample.
We consider both of the points representing the $z\sim6$ quasar space density considering the small number statistics and the small redshift difference.

Despite the small number of our sample, we additionally calculated the binned space densities in the three survey fields where the quasars are identified: EGS ($i'_1$), EGS ($i'_2$), and SA22($i'_1$).
The results are given in the bottom three rows in Table \ref{tbl:binqlf}, which are higher than the one for the total survey area.
We discuss this in the following section.

\section{DISCUSSION\label{sec:discussion}}

\begin{figure*}
\centering
\epsscale{1.2}
\plotone{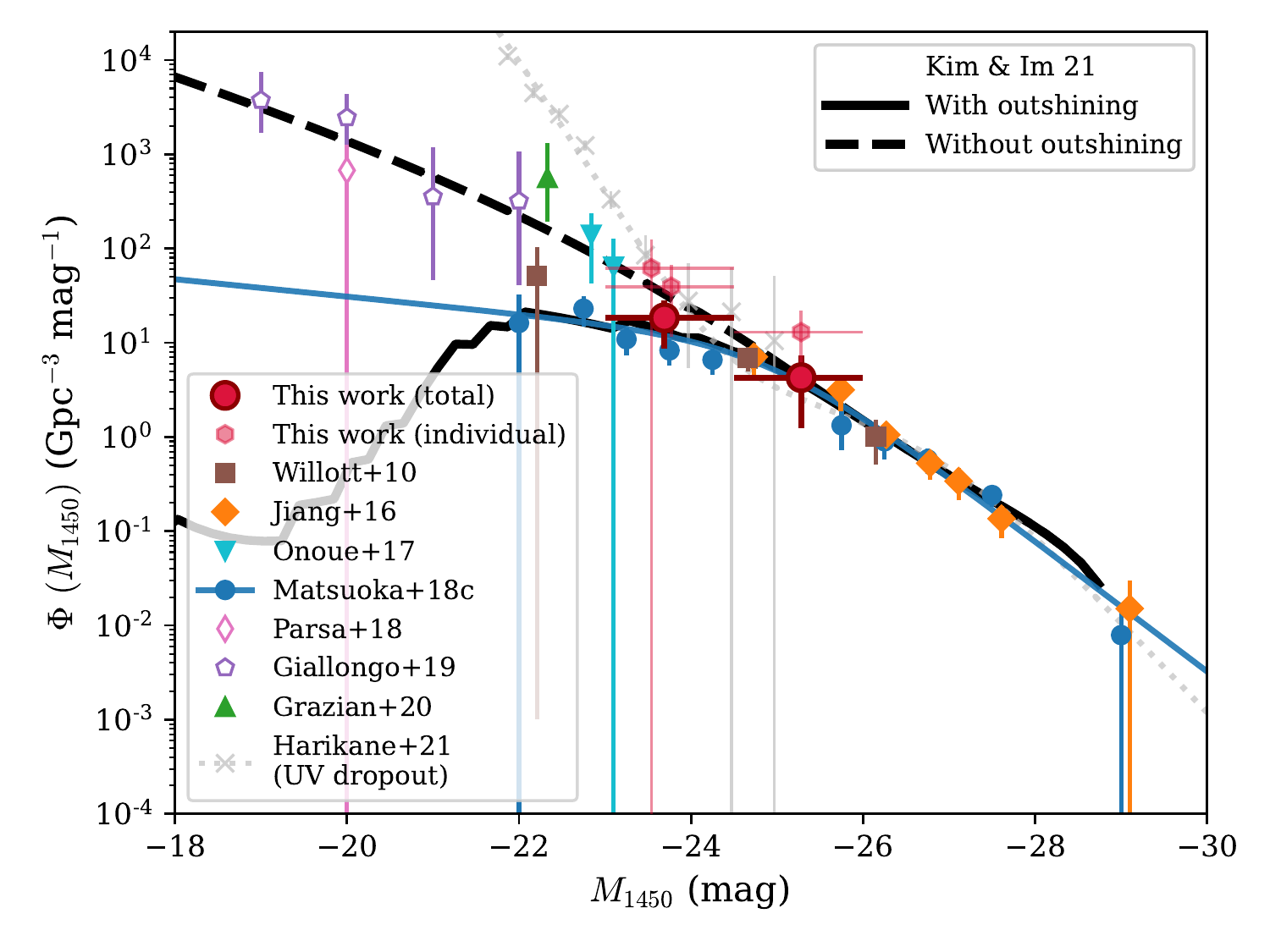}
\caption{
Quasar space densities at $z\sim6$.
The red circles represent our results from the total survey area, while the red hexagons are from the three individual fields where high-redshift quasars are identified.
The other symbols represent those in the literature \citep{Willott10b,Jiang16,Onoue17,Matsuoka18c,Parsa18,Giallongo19,Grazian20}.
The filled (open) symbols are from the surveys based on rest-UV photometry (X-ray detection).
The blue line shows the parametric LF of quasar \citep{Matsuoka18c}.
The black solid (dashed) line represents the quasar LF model with (without) the outshining effect \citep{Kim21}.
The parametric LF of UV dropout objects (or AGN+galaxy; \citealt{Harikane21}) is auxiliary shown as the gray crosses with dotted line.
\label{fig:qlf}}
\end{figure*}

\begin{deluxetable}{lccccc}
\tabletypesize{\scriptsize}
\tablecaption{Binned Space Density of IMS $z\sim6$ quasars\label{tbl:binqlf}}
\tablehead{
\colhead{Field} & \colhead{$M_{1450}$}  & \colhead{$\Delta M_{1450}$} & \colhead{$N_{\rm bin}$} & \colhead{$V_{a}$}  & \colhead{$\Phibin$}\\
\colhead{} & \colhead{(mag)} & \colhead{(mag)} &  & \colhead{ (Gpc$^{3}$)} & \colhead{ (Gpc$^{-3}$ mag$^{-1}$)}
}
\startdata
Total & $-25.28$ & 1.5 & 2 & 0.31 & $4.27\pm3.02$ \\ 
 & $-23.69$ & 1.5 & 3 & 0.14 & $13.8\pm8.0$ \\
\hline
EGS ($i'_1$)  & $-25.28$ & 1.5 & 2 & 0.10 & $13.0\pm9.2$\\
EGS ($i'_2$)  & $-23.54$ & 1.5 & 1 & 0.01 & $61.9\pm61.9$\\
SA22 ($i'_1$)& $-23.77$ & 1.5 & 2 & 0.03 & $39.3\pm27.8$\\
\enddata
\end{deluxetable}

In Figure \ref{fig:qlf}, we compare our results with those from the literature \citep{Willott10b,Jiang16,Matsuoka18c,Giallongo19,Grazian20}, after the correction for the cosmological parameters.
For the results from faint X-ray AGNs at $z\sim5.5$ \citep{Parsa18,Giallongo19,Grazian20}, we adopted the density shift to $z=6$ using the density scaling factor of $10^{-0.72\Delta z}$ at $z=5$--6 \citep{Jiang16}.
Note that \cite{Matsuoka18c} derived their space densities including the samples of \cite{Willott10b} and \cite{Jiang16}.
We also show \cite{Matsuoka18c}'s LF in a double-power law function (blue solid line).
Our space densities from the total survey area (red circles) are broadly consistent with those from the previous large surveys for UV quasars \citep{Willott10b,Jiang16,Matsuoka18c}, despite large errors with the small number statistics.
In the intermediate magnitude range of $-24<M_{1450}<-22$ in question, our result shows the suppressed space density in line with the recent quasar LF of \cite{Matsuoka18c}.
Therefore, our result reinforces the suggestion that quasars are not the main contributor to the reionizing process at $z\sim6$ (e.g., \citealt{Ricci17,Dayal20,Jiang22}), disfavoring the AGN-dominant scenario (e.g., \citealt{Giallongo15,Madau15}).

Our main result, however, is somewhat different from the space densities of \cite{Onoue17} and \cite{Grazian20}, both from the AGNs identified by rest-UV spectroscopy, favoring a continuous increase in space density from bright to faint AGN populations.
But we here point out that the fundamental limitation of these two studies is their small survey areas (6.5 and 0.15 deg$^{2}$, respectively) and corresponding small $V_{a}$, which could result in the overestimated space densities.
For instance, we show the space densities from our three individual fields where high-redshift quasars are discovered (red hexagons): EGS ($i'_1$), EGS ($i'_2$), and SA22 ($i'_1$).
As in the two studies, the results are from one or two quasars in the small survey area (29.2, 5.2, and 16.7 deg$^{2}$), which may give higher space densities than those from our total survey area.

\cite{Grazian20} suggest that their higher space density is due to the stringent color selection criteria of the other studies (e.g., $i'_{\rm HSC}-z'_{\rm HSC}>2$ in \citealt{Matsuoka18c}), while the two quasars they used (GDN 3333 and GDS 3073) have moderate $(i'-z')$-matched colors; 0.12 and 0.69, respectively.
But the two quasars are at $z=5.2$ and 5.6, respectively, so it would be better to check their $r'-i'$ colors instead to see whether they can be selected by the traditional color selection.
From the CANDELS catalogs \citep{Guo13,Barro19}, we found that their ($r'-i'$)-matched colors\footnote{F606W$-$F775W color in the \textit{Hubble Space Telescope}/Advanced Camera for Surveys (HST/ACS) filter system.} are 2.11 and 3.03, respectively.
These are red enough to be selected as a quasar candidate with a color selection (e.g., $r'-i'>1.2$ for $z>5$ quasars; \citealt{Kim20}), so the strict color selection cannot solely explain the discrepancy clearly.

From a different point of view, such high densities can be explained with the recent quasar LF model by \cite{Kim21}.
This model is based on the empirical scaling relations of dark matter halos, galaxies, and black holes, while the key idea is that an AGN outshining its host galaxy can be observed as a point-source-like quasar.
In Figure \ref{fig:qlf}, we show the model with/without the outshining effect as the black solid/dashed lines, respectively.
Note that we show the models including the gravitational lensing effect of \cite{Pacucci20}.
These models suggest that the discrepancy in space density between rest-UV quasars (e.g., \citealt{Matsuoka18c}; this work) and faint X-ray AGNs from some previous works (e.g., \citealt{Giallongo19}) can be explained if a large fraction of AGNs become dimmer in UV than its host galaxy.

The high space densities of \cite{Onoue17} and \cite{Grazian20} are in line with the model without the outshining effect.
It is worth noting that the AGNs used in these studies have distinct properties from typical bright quasars.
For example, ELAIS109100446, one of the two AGNs in \cite{Onoue17}, has only a narrow $\lya$ line without any other emission lines on its rest-UV spectrum, which indicates that it could be a $\lya$ emitter galaxy \citep{Kashikawa15}.
GDN 3333 in \cite{Grazian20} shows similar features on its spectrum \citep{Barger02}, while it is classified as an AGN with strong X-ray detection \citep{Alexander03,Giallongo19}.
GDS 3073, another quasar in \cite{Grazian20}, is likely a Seyfert galaxy based on the morphological decomposition, while it has no X-ray detection.
Taken together, their observational properties appear to be a mixture of AGNs and UV-bright galaxies, so their presence is consistent with the framework of \cite{Kim21}.

It has been recently claimed that the $M_{1450}$ boundary between AGN and star-forming galaxy (represented by Lyman-break galaxy) is blurred, i.e., the AGN fraction of UV sources changes smoothly at $-24\lesssim M_{1450}\lesssim-22$ \citep{Adams20,Bowler21,Harikane21}.
In Figure \ref{fig:qlf}, we show the LF of UV dropout objects at $z\sim6$ (gray crosses; \citealt{Harikane21}).
Compared to their best-fit result (gray dotted line), our faint bin gives an AGN fraction of $\sim30\%$ at $M_{1450}=-23.7$ mag, which is lower than those at lower redshifts (e.g., $\sim80\%$ at $z\sim4$; \citealt{Bowler21}).
This is naturally explained by the more dramatic changes in AGN numbers between $4\lesssim z\lesssim7$ \citep{Akiyama18,Matsuoka18c,Wang19,Kim20,Niida20} than galaxy numbers in UV \citep{Song16,Ono18,Behroozi19,Harikane21}.

\section{SUMMARY}

In this work, we present the final result of the IMS $z\sim6$ quasar survey.
Over the $86$ deg$^{2}$ sky area of CFHTLS-IMS overlap regions, 25 candidates satisfying the traditional color selection criteria were picked up.
We additionally applied the AICc selection based on the SED fitting, resulting in the seven credible candidates.
While three of them are known $z\sim6$ quasars, our follow-up spectroscopy for the remaining candidates leads us to discover two new $z\sim6$ quasars.
Such a high success rate (5/7) proves that our new approach with the AICc method allows us to find plausible candidates efficiently.
With the complete sample of five quasars, we estimated the quasar space density down to $M_{1450}=-23.5$ mag at $z\sim6$; $\Phibin=4.3$ and 14 Gpc$^{-3}$ mag$^{-1}$ at $M_{1450}=-25.3$ and $-23.7$ mag, respectively.
These low numbers are consistent with the recent estimates from other large surveys, which endorses the minor role of quasars in the ionizing process in the reionization era.

\acknowledgments

We thank Marios Karouzos and Jueun Hong for their valuable contribution to IMS.
We also thank the anonymous referee for suggestions that improved the quality of the draft.
This work was supported by the National Key R\&D Program of China (2016YFA0400703) and the National Research Foundation of Korea (NRF) grant funded by the Korean government (MSIT) (No. 2020R1A2C3011091, 2021M3F7A1084525).
Y. K. acknowledges the support from the China Postdoc Science General (2020M670022) and Special (2020T130018) Grants funded by the China Postdoctoral Science Foundation, and was supported by the National Research Foundation of Korea (NRF) grant funded by the Korean government (MSIT) (No. 2021R1C1C2091550).
M. K. was supported by the National Research Foundation of Korea (NRF) (No. 2022R1A4A3031306).
S. S. acknowledges support by Basic Science research Program through the National Research Program through the National Research Foundation of Korea (NRF) funded by the Ministry of Education (No. 2020R1A6A3A13069198).
M. H. acknowledges the support from the Korea Astronomy and Space Science Institute grant funded by Korea government (MSIT) (No. 2022183005).
B. L. acknowledges the support from the Korea Astronomy and Space Science Institute grant funded by the Korea government (MSIT) (Project No. 2022-1-840-05).
H. D. J. was supported by the National Research Foundation of Korea (NRF) grant 2022R1C1C2013543 funded by the Ministry of Science and ICT (MSIT) of Korea.
D. K. was supported by the National Research Foundation of Korea (NRF) grant funded by the Korea government (MSIT) (No. 2021R1C1C1013580 and 2022R1A4A3031306).
S. L. acknowledges support from a National Research Foundation of Korea (NRF) grant (2020R1I1A1A01060310) funded by the Korean government (MIST).
Y. C. T. acknowledges the support from the Basic Science Research Program through the 
National Research Foundation of Korea (NRF) funded by the Ministry of Education (No. 2021R1A6A3A14044070).

This research uses data obtained through the Telescope Access Program (TAP) (PID: CTAP2020-B0043, CTAP2021-A0032). The data from the former program is not included in this paper, because of the its condition.
Observations obtained with the Hale Telescope at Palomar Observatory were obtained as part of an agreement between the National Astronomical Observations, Chinese Academy of Sciences, and the California Institute of Technology.

This work was supported by K-GMT Science Program (PID: GS-2020A-Q-219) of the Korean Astronomy and Space Science Institute (KASI).
Based on observations obtained at the international Gemini Observatory, a program of NSF’s NOIRLab, which is managed by the Association of Universities for Research in Astronomy (AURA) under a cooperative agreement with the National Science Foundation on behalf of the Gemini Observatory partnership: the National Science Foundation (United States), National Research Council (Canada), Agencia Nacional de Investigaci\'{o}n y Desarrollo (Chile), Ministerio de Ciencia, Tecnolog\'{i}a e Innovaci\'{o}n (Argentina), Minist\'{e}rio da Ci\^{e}ncia, Tecnologia, Inova\c{c}\~{o}es e Comunica\c{c}\~{o}es (Brazil), and Korea Astronomy and Space Science Institute (Republic of Korea).

The Pan-STARRS1 Surveys (PS1) and the PS1 public science archive have been made possible through contributions by the Institute for Astronomy, the University of Hawaii, the Pan-STARRS Project Office, the Max-Planck Society and its participating institutes, the Max Planck Institute for Astronomy, Heidelberg and the Max Planck Institute for Extraterrestrial Physics, Garching, The Johns Hopkins University, Durham University, the University of Edinburgh, the Queen's University Belfast, the Harvard-Smithsonian Center for Astrophysics, the Las Cumbres Observatory Global Telescope Network Incorporated, the National Central University of Taiwan, the Space Telescope Science Institute, the National Aeronautics and Space Administration under Grant No. NNX08AR22G issued through the Planetary Science Division of the NASA Science Mission Directorate, the National Science Foundation Grant No. AST-1238877, the University of Maryland, Eotvos Lorand University (ELTE), the Los Alamos National Laboratory, and the Gordon and Betty Moore Foundation.

This publication makes use of data products from the Two Micron All Sky Survey, which is a joint project of the University of Massachusetts and the Infrared Processing and Analysis Center/California Institute of Technology, funded by the National Aeronautics and Space Administration and the National Science Foundation.

\facilities{CFHT (MegaCam), Hale (DBSP), Gemini:South (GMOS-S), UKIRT (WFCAM)}

\software{
PSFEx \citep{Bertin11}, PypeIt \citep{Prochaska20a,Prochaska20b}, SExtractor \citep{Bertin96}, SWarp \citep{Bertin10}}

\appendix

\section{Quasar Candidate Selection without $J$-band detection\label{sec:nojdet}}

The fifth criterion of our initial color selection ($J<J_{5\sigma}$ in Section \ref{sec:cs}) allows us to select only the sources with significant $J$-band detections.
But, we may miss many quasars due to this criterion. 
For example, a $z=6.0$ quasar at $z'=23.0$ mag is expected to have $J\sim22.8$ mag, inferred from the quasar track in Figure \ref{fig:ccd}, which is beyond the $J$-band $5\sigma$ depth of many of the survey areas (especially for those in CFHTLS-W2 and EGS fields; see Figure \ref{fig:histdepth}).

\begin{figure}
\centering
\epsscale{1}
\plotone{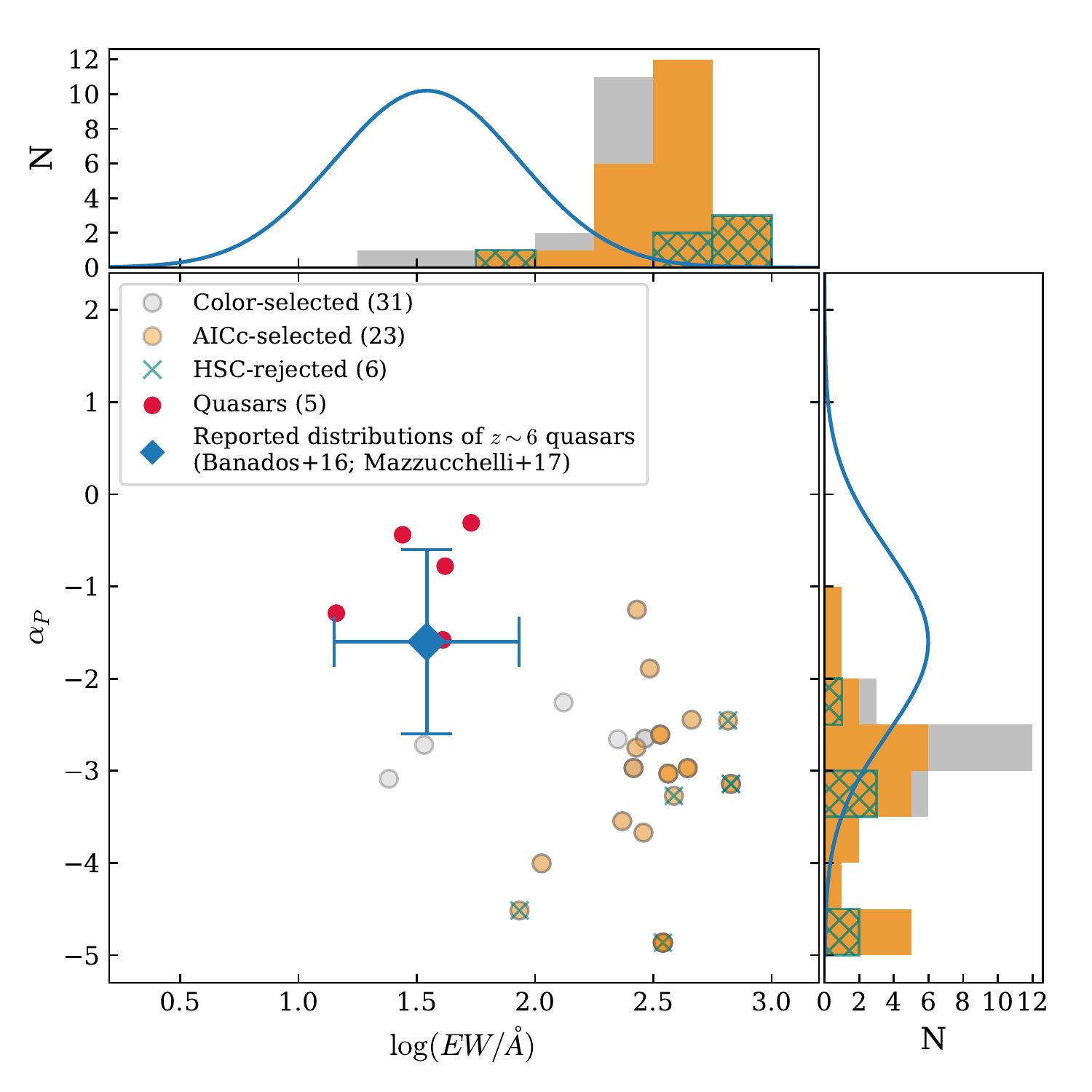}
\caption{
EW and $\alpha_{P}$ distributions of the candidates without $J$-band detection.
The gray circles are the color-selected candidates, while the orange ones are the ones satisfying the AICc criterion ($w_{q}>0.99$).
Several candidates matched to the same quasar model were overlaid, so they appear to have low transparency on the diagram.
The candidates covered by the HSC survey are marked by teal crosses.
The five quasars in this work are shown as red circles.
The blue diamond (and blue lines on the histograms) represents the reported normal distribution of $z\sim6$ quasars: $\log EW=1.542\pm0.391$ \citep{Banados16} and $\alpha_{P}=-1.6\pm1.0$ \citep{Mazzucchelli17}, in a form of mean$\pm$standard deviation.
Note that the correlation between the reported EW and $\alpha_{P}$ is not reflected in this figure.
\label{fig:nojdet}}
\end{figure}

We performed the same selection process as in the main text except for the fifth criterion.
Additional 1563 sources were color-selected, and 899 sources remain after the automatic process to reject sources on the bad pixels.
Visual inspection of the images of the remaining sources showed that most of them are cosmic rays and diffraction spikes.
So, we finally have 31 sources after the visual inspection.
Interestingly, 23 of 31 objects have $w_q>0.99$ with our AICc method, meaning that they are likely to be high-redshift quasars rather than late-type stars.
However, we note that their best-fit models correspond to extreme cases.
Figure \ref{fig:nojdet} shows the EW and $\alpha_{P}$ distributions of the candidates without $J$-band detection.
Compared to the five quasars in this work (red circles; Table \ref{tbl:sample}) and the reported distributions of $z\sim6$ quasars (blue diamond; \citealt{Banados16,Mazzucchelli17}), they have much higher EW values and steeper continuum slope, so they are unlikely to be high-redshift quasars.

Recent high-redshift quasar survey studies with deep HSC data have discovered some $z\sim6$ quasars with strong but narrow $\lya$ line (EW$>200~\rm\AA$; \citealt{Matsuoka18a,Matsuoka18b,Matsuoka19a,Matsuoka22}), but most of them are fainter than our survey limit ($M_{1450}>-23.5$ mag).
\cite{Matsuoka22} also showed that such faint quasars tend to have a strong $\lya$ line (or a high EW compared to continuum).
Note that the EW values are estimated with the fixed continuum slope of $\alpha_{P}=-1.5$ in those studies, so our EW values will increase if we assume the same condition.
In addition, we cross-checked the reliability of the 31 candidates with deep HSC images.
Only six out of them are located in the HSC PDR3 survey area (teal crosses in Figure \ref{fig:nojdet}), but five of them have no matched sources even though they are bright enough to be detected in the HSC images.
This means that such sources in our survey may be non-celestial bodies or artificial objects just detected in the $z'$-band images.
Even for the matched one ($\log{\rm EW}=1.9$ \& $\alpha_{P}=-4.5$ ), it has a very blue color of $i'_{\rm HSC}-z'_{\rm HSC}=0.38$.
Therefore, we conclude that these additional candidates without $J$-band detection are unlikely to be real high-redshift quasars.

\end{document}